\documentclass[12pt]{article}
\usepackage{amsfonts}
\usepackage{latexsym}
\usepackage{amsmath}
\usepackage{amssymb}
\usepackage{amssymb}

\newcommand{\newsection}{    
\setcounter{equation}{0}\section}
\def\appendix#1{\addtocounter{section}{1}\setcounter{equation}{0}
\renewcommand{\thesection}{\Alph{section}}
\section*{Appendix \thesection\protect\indent \parbox[t]{11.15cm}{#1}}
\addcontentsline{toc}{section}{Appendix \thesection\ \ \ #1}}

\newcommand{\be}{\begin{eqnarray}}
\newcommand{\ee}{\end{eqnarray}}
\newcommand{\bea}{\begin{eqnarray}}
\newcommand{\eea}{\end{eqnarray}}
\newcommand{\ba}{\begin{array}}
\newcommand{\ea}{\end{array}}
\newcommand{\nn}{\nonumber \\}

\def \la {\label}

\def\a{\alpha}
\def\b{\beta}

\def\bbe{{\bf{e}}}
\font\mybb=msbm10 at 11pt

\def\bb#1{\hbox{\mybb#1}}

\def\bZ {\bb{Z}}
\def\bR {\bb{R}}

\def\bH {\bb{H}}

\def\bP {\bb{P}}

\def\ui {{\underline {i}}}
\def\uj {{\underline {j}}}
\def\uk {{\underline {k}}}
\def\uell {{\underline {\ell}}}

\begin{document}
\begin{titlepage}
\begin{center}
\vspace{5.0cm}


\vspace{3.0cm} {\Large \bf Heterotic horizons,  Monge-Amp\`ere equation and del Pezzo surfaces}

\vspace{3.0cm}


{}\vspace{2.0cm}
 {\large
J.~Gutowski and  G.~Papadopoulos
 }

{}

\vspace{1.0cm}
Department of Mathematics\\
King's College London\\
Strand\\
London WC2R 2LS, UK\\

\end{center}
{}
\vskip 3.0 cm
\begin{abstract}

Heterotic horizons  preserving 4 supersymmetries have sections which are $T^2$ fibrations over 6-dimensional conformally balanced Hermitian manifolds.
We give new examples of  horizons with sections $S^3\times S^3\times T^2$ and $SU(3)$.
We then examine the  heterotic horizons which are $T^4$ fibrations over a K\"ahler 4-dimensional
manifold. We prove that the solutions depend on 6 functions which are determined by a  non-linear differential system of 6 equations
  that
include the Monge-Amp\'ere  equation. We show that this system has an explicit solution
for the K\"ahler manifold $S^2\times S^2$.
We also demonstrate that there is an associated cohomological system which has solutions on
del Pezzo surfaces. We
raise the question of whether for every solution of the cohomological problem
there is a solution of the differential system, and so a new heterotic horizon.  The horizon sections
have topologies which include $((k-1) S^2\times S^4\# k (S^3\times S^3))\times T^2$  indicating the existence of exotic black holes.
We also find an example of a horizon section which  gives rise to two different near horizon geometries.

\end{abstract}
\end{titlepage}

\setcounter{section}{0}
\setcounter{subsection}{0}


\newsection{Introduction}

In the past few years much work has been done to understand the topology and geometry of higher dimensional black holes following earlier
work in four dimensions \cite{israel}-\cite{robinson}.
It has been realized that the four dimensional uniqueness theorems fail to hold in higher dimensions, and that the horizon sections can have many different topologies,
including $S^n\times S^m$, $n\geq 0, m\geq 1$ \cite{reallbh}-\cite{kl}.

More recently, the near horizon geometry of supersymmetric heterotic black holes has been investigated \cite{hh} utilizing
the solution of the Killing spinor equations of heterotic supergravity in \cite{het1}. It was found that
the heterotic horizons are either products $AdS_3\times X$ for a suitable manifold $X$ or $(AdS_3\times Y)/S^1$, where $AdS_3$ twists over a base space
with
a $U(1)$ connection. The heterotic horizons preserving  8 supersymmetries are isometric to $AdS_3\times S^3\times T^4$
or $AdS_3\times S^3\times K_3$ with constant dilaton and have horizon section $S^1\times S^3\times T^4$ or $S^1\times S^3\times K_3$, respectively.

In this paper, we shall give new heterotic horizons which preserve $N=4$ supersymmetries. Moreover, we shall provide evidence
that heterotic black holes can have increasingly involved horizon topologies which are distinct from those expected from
lifting 4- and 5-dimensional black holes to 10 dimensions. It is known that
$N=4$ horizon sections ${\cal S}$ are holomorphic $T^2$ fibrations over a complex conformally balanced 6-dimensional
manifold $B$. First, we shall reorganize the differential conditions which arise from supersymmetry in a
form which is amenable to a cohomological analysis. Then we shall demonstrate that the differential system has solutions
provided that certain conditions on the cohomology of $B$ are satisfied.
 We  present explicit heterotic horizon
solutions with  $B=SU(3)/T^2$ equipped with the balanced Hermitian structure and $B=(S^3\times S^3)/S^1\times S^1$. The horizon sections
are ${\cal S}=SU(3)$ and ${\cal S}=S^3\times S^3\times T^2$, and the spacetime is $M=\big(SL(2,\bR)\times SU(3)\big)/U(1)$
and $M=AdS_3\times S^3\times S^3\times S^1$, respectively. In the former case, $AdS_3$ twists over $B$ with a $U(1)$ connection. In the latter case,
the radius of $AdS_3$ is twice that of $S^3$.  The half supersymmetric horizon $AdS_3\times S^3\times K_3$ is also of this type with $B=\bP^1\times K_3$
K\"ahler and so Hermitian and balanced.

To construct  more examples of heterotic horizons, one can take $B$ to be a holomorphic $T^2$-fibration over a 4-dimensional K\"ahler manifold $X$.
Such complex manifolds with skew-symmetric torsion and holonomy contained in $SU(3)$ have been considered before in  \cite{sethi, gold, grant, yaufu}.
For such $B$, the
horizon section ${\cal S}$ is a $T^4$ fibration over $X$. We show that the resulting field equations and supersymmetry
conditions lead to a non-linear system of 6 differential equations for 6 functions\footnote{The system we derive is different from that of
\cite{yaufu} as in our case the torsion is closed on the horizon section ${\cal S}$.}.
The 6 functions include the dilaton, the deformation of the K\"ahler metric of $X$ within its K\"ahler class as well as
the deformations of connections of holomorphic $U(1)$ bundles within their Chern classes. The deformation of the metric
within its K\"ahler class leads to a complex Monge-Amp\`ere equation similar to that which appears in  Yau's proof of the Calabi conjecture.
The system also
includes a conformally rescaled Hermitian-Einstein equation. However, the six differential equations do not separate. As in the general case,
the differential equations give rise to some cohomological conditions for classes in $X$. These are necessary for the differential system
to have a solution. We raise the question of whether {\it for every solution of the cohomological problem there is also
a solution of the differential system}.

The non-linear differential  system of equations has solutions.  One explicit solution is
$X=\bP^1\times \bP^1$, $B=\bP^1\times \bP^1\times T^2$ and ${\cal S}=S^3\times S^3\times T^2$. The heterotic horizon spacetime
is $M=\big(SL(2, \bR)\times S^3\times S^3\times T^2\big)/S^1$. Another example is the half supersymmetric
horizon $AdS_3\times S^3\times T^4$ with $X=\bP^1\times T^2$, and either $B=\bP^1\times T^2\times T^2$ or $B=S^3\times T^3$, and ${\cal S}=S^3\times T^5$.
One of the consequences of the explicit examples we have constructed is that the same horizon section ${\cal S}=S^3\times S^3\times T^2$
gives rise to two different near horizon geometries $M=AdS_3\times S^3\times S^3\times S^1$ and $M=\big(SL(2, \bR)\times S^3\times S^3\times T^2\big)/S^1$.
As far as we know, this is the first time that such a phenomenon  has been observed. As such it represents an additional difficulty in the classification
of black holes in higher dimensions as well as in the investigation of their thermodynamical properties.

To give evidence that there are more heterotic horizons, we
 demonstrate that the associated cohomological conditions have solutions for $X$ a del Pezzo surface. Considering the del Pezzo surfaces
 as $\bP^2$ blown up at $k<9$ points, we find that if $k$ is odd the cohomological conditions are met provided that the K\"ahler class is identified with
 the anti-canonical class. If $k$ is even, then a different choice for a K\"ahler class has to be made. There are solutions for all del Pezzo surfaces.
 We also investigate the topology of the associated heterotic horizons. For this we  compute the de Rham cohomology
 of ${\cal S}$. We find that in some cases the horizon sections have the same de Rham cohomology as that of
$\big((k-1)(S^2\times S^4)\# k(S^3\times S^3)\big) \times T^2$, and  under an additional assumption they are diffeomorphic  to them.
We also point out that the cohomology of these black hole horizons, and in particular some intersection matrices,
are related to exceptional groups. This is inherited from the relation of the intersection matrix of the second cohomology of del Pezzo
surfaces to the Cartan matrix of the exceptional groups. Our cohomological conditions do not have a solution on $\bP^2$ blown up
at more than 9 points. However this restriction can be removed if the Bianchi identity
of the 3-form field strength is modified either by adding point sources or by taking into account the heterotic anomaly.

We also explore the possibility of extending our near horizon geometries to full black hole solutions. One may expect that
there is a continuous interpolation between a horizon section and a section of the asymptotic geometry of a black hole, ie the two sections
are cobordant. We  argue that in the presence of fermions, and in particular supersymmetry, the two sections must represent the same class
in the spin co-bordism ring $\Omega^{\mathfrak{spin}}_*$. Similar tests have been proposed elsewhere \cite{kl} using oriented cobordism. We find that most of our horizons can be associated with asymptotically flat or $AdS$
black holes.

Since for all our solutions the 3-form field strength is closed, they can also be interpreted as solutions IIA, IIB and 11-dimensional
supergravity. Lifting our heterotic solutions to 11-dimensional supergravity and then reducing them in a different direction, we show that
new solutions can be constructed in IIA supergravity which have an $AdS_2$ component and RR fluxes. These can also be further T-dualized
to find new solutions in IIB supergravity. Interpreting these new solutions as near horizon geometries, we provide evidence
that both IIA and IIB supergravity admit black hole solutions with non-trivial topology and with  all form fluxes
non-vanishing.

In the cases that have been investigated so far, the near horizon supersymmetric solutions also arise
as near brane geometries. For the $1/2$ supersymmetric near horizon heterotic geometries
this has been demonstrated in \cite{hh}. This is also  the case for  the
$1/4$ supersymmetric  $AdS_3\times S^3\times S^3\times S^1$ solution. It turns out
that this is
the near brane geometry of two  5-branes with a localized string superposition. It is likely that the rest
of the solutions have a near brane interpretation.

This paper is organized as follows. In section two, we use the results of \cite{hh} to reconstruct the geometry
of the horizon $M$ and that of the horizon section ${\cal S}$ from geometric data on $B$ and those of the toric fibration over $B$.
We also summarize and extend some results of \cite{gold} on the relation between Hermitian conformally balanced manifolds
and toric fibrations. In section 3, we give the differential system needed to construct heterotic horizon sections
as $T^2$ fibrations over a conformally balanced Hermitian 6-dimensional manifold. We present an explicit solution with horizon section
$SU(3)$. In section 4, we  construct heterotic horizon sections
as $T^4$ fibrations over K\"ahler 4-dimensional manifolds $X$. We show that this leads to  a differential system of 6 non-linear equations
for 6 functions and has a cohomological analogue in the cohomology ring of $X$. We give an explicit example that solves
the differential system. In section 5, we  find  solutions of the cohomological problem on del Pezzo surfaces. In section 6,
we find brane configurations with near brane geometries similar to those that arise as near horizon geometries. In section 7,
we explore our solutions in type II and 11-dimensional supergravities. In section 8, we use spin cobordism to
provide evidence that our near horizon geometries can be extended to black hole solutions and in section 9, we give our conclusions.
In appendix A, we compute the cohomology of horizon sections, and in appendix B we generalized the differential system that arises
in the context of heterotic horizons.

\newsection{Geometric conditions}

Heterotic horizons which preserve 4 supersymmetries, and their associated sections, are fibrations over a 6-dimensional manifold $B^6$. In
particular, the horizon spacetime $M$ is a $SL(2,\bR)\times U(1)$ fibration while the section ${\cal S}$ is a $T^2$ fibration.
The fibre directions twist over  $B^6$ with a  non-trivial connection.
In what follows, we shall reconstruct both $M$ and ${\cal S}$ from geometric data given on $B^6$ and on the twisting of the fibration.

\subsection{Reconstruction of the horizon sections}

The geometry of a  horizon spacetime $M$ is completely determined in terms of the geometric data on the horizon section ${\cal S}$. Because of this,
we first reconstruct ${\cal S}$ as a $T^2$ fibration over $B^6$. Let
  $ds_{(6)}^2$ be the metric and $H_{(6)}$ be the torsion of $B^6$.
The metric and torsion of ${\cal S}$ can be written as
\bea
d\tilde s_{(8)}^2= k^{-2}h\otimes h+k^{-2}\ell\otimes\ell+ds_{(6)}^2~,~~~\tilde H_{(8)}=k^{-2} h\wedge dh+k^{-2}\ell\wedge d\ell+H_{(6)}~,
\la{8data}
\eea
respectively, where $h$ and $\ell$ are 1-forms  on ${\cal S}$ which are interpreted as principal bundle connections associated with the
fibration, and $k^2$ is the constant length of $h$ and $\ell$. Both $d\tilde s_{(8)}^2$ and $\tilde H_{(8)}$
are invariant under the rotations of the $T^2$ fibre generated by the vector fields $\xi$ and $\eta$ dual to $h$ and $\ell$, respectively,
relative to the metric\footnote{One could write a more general metric on ${\cal S}$ by introducing a non-trivial constant metric on the fibre directions.
The choice we have made suffices for our purpose.} $d\tilde s_{(8)}^2$. In particular, one has
\bea
h=d\tau+\alpha_i e^i~,~~~\ell=d\sigma+\beta_i e^i
\eea
where $0\leq \tau, \sigma\leq 2\pi$ are the coordinates of $T^2$, $\alpha$ and $\beta$ are connections on $B^6$ and $(e^i)$ is a local frame,
$ds_{(6)}^2=\delta_{ij} e^i e^j$. The components of the metric and torsion depend only on the coordinates of $B^6$.

Supersymmetry restricts both the connection $h$ and $\ell$, and the geometry of the base space $B^6$ as follows.
$B^6$ is a Hermitian manifold with Hermitian form $\omega_{(6)}$ which is compatible with the metric connection, $\hat{\nabla}_{(6)}$,
 with skew-symmetric torsion, $H_{(6)}$, ie
\bea
\hat{\nabla}_{(6)}\omega_{(6)}=0~.
\eea
This is equivalent to setting
\bea
H_{(6)}=-i_{I_{(6)}} d\omega_{(6)}~,~~~dH_{(6)}\not=0~,
\la{6hi}
\eea
where $I_{(6)}$ is the complex structure on $B^6$.  Moreover, it is required that $B^6$ is conformally balanced. This means that the Lee form
of $B^6$ is
\bea
\theta_{\omega_{(6)}}=2d\Phi~,
\eea
where $\Phi$ is the dilaton that depends only on the coordinates of $B^6$. This summarizes the conditions on $B^6$.

The restriction on the twisting of the $T^2$ fibration over $B^6$ is made by putting appropriate conditions on the
connections $h$ and $\ell$. In particular, $N=4$ supersymmetry requires that
\bea
dh^{2,0}=d\ell^{2,0}=0~,~~~dh_{ij} \omega^{ij}_{(6)}=0~,~~~d\ell_{ij} \omega^{ij}_{(6)}=-2 k^2~.
\eea
This means that the curvature of the torus fibration is (1,1) with respect to the complex structure of $B^6$ and
in addition one of the components of the connection is traceless while the other has constant trace.

The first two conditions on the curvature of the fibration can be solved by requiring that ${\cal S}$ is a holomorphic
fibration. It is straightforward to see that the Hermitian form
\bea
\omega_{(8)}= {1\over k^2} h\wedge \ell+\omega_{(6)}~,
\eea
gives rise to an integrable complex structure on ${\cal S}$.

Collecting the above data, one finds that
\bea
\hat{\tilde \nabla}_{(8)}\omega_{(8)}=0~,~~~\hat{\tilde \nabla}_{(8)}h=\hat{\tilde \nabla}_{(8)}\ell=0~,
\la{parcon}
\eea
where $\hat{\tilde \nabla}_{(8)}$ is the metric connection on ${\cal S}$ with skew-symmetric torsion $H_{(8)}$. Consistency
therefore requires that
\bea
\tilde H_{(8)}=-i_{I_{(8)}} d\omega_{(8)}~,
\eea
which is equivalent to (\ref{8data}) and (\ref{6hi}).
So far the data imply that  the holonomy of $\hat{\tilde \nabla}_{(8)}$ is contained in $U(3)$, ${\rm hol}(\hat{\tilde \nabla}_{(8)})\subseteq U(3)$.

Furthermore, supersymmetry requires that the holonomy of $\hat{\tilde \nabla}_{(8)}$ must be contained in $SU(3)$. One way to enforce this
is to require that the Ricci form of the $\hat{\tilde \nabla}_{(8)}$ connection vanishes
\bea
\hat\rho_{(8)}={1\over4}  (\hat{\tilde R}_{(8)})_{\uk\uell,}{}^\ui{}_\uj I_{(8)}^\uj{}_\ui e^\uk\wedge e^\uell=0~,
\eea
where $\uk, \uell$ are ${\cal S}$ frame indices and $\hat{\tilde R}_{(8)}$ is the curvature of the connection $\hat{\tilde \nabla}_{(8)}$.
This requirement gives several conditions.  The only one which is independent  from those that
have already been stated is
\bea
 (\hat\rho_{(6)})_{ij}-  d\ell_{ij}=0~,
\eea
where $\hat\rho_{(6)}$ is the Ricci form of the $\hat{\nabla}_{(6)}$ connection. If $B^6$ is simply connected, this and (\ref{parcon}) are
necessary and
sufficient conditions for ${\rm hol}(\hat{\tilde \nabla}_{(8)})\subseteq SU(3)$. In particular, this implies that there is a (3,0) form $\chi$
such that
\bea
\hat{\tilde \nabla}_{(8)}\chi=0~,~~i_\xi \chi=i_\eta \chi=0~,
\eea
as stated in \cite{hh}. The geometries of $B^6$ and ${\cal S}$ are summarized in table 1.

\begin{table}[ht]
 \begin{center}
\begin{tabular}{|c|c|c|}
\hline
$\mathrm{Geometry}$&$B^6$&${\cal S}$\\
\hline
\hline
$\mathrm {Hermitian}$&$\mathrm {yes}$&$\mathrm {yes}$ \\
 \hline
$\mathrm {Conformally ~balanced}$&$\mathrm {yes}$&$\mathrm {no}$
\\
$\theta=2d\Phi$&&\\
\hline
$\mathrm{hol}(\hat\nabla)\subseteq SU(3)$&$\mathrm {no}$&$\mathrm {yes}$\\
\hline
$\mathrm{hol}(\hat\nabla)\subseteq U(3)$&$\mathrm {yes}$&$\mathrm{ no}$\\
\hline
$dH_{(n)}=0$&$\mathrm {no}$&$\mathrm{ yes}$\\
\hline
\end{tabular}
\end{center}
\caption{\small
The geometry of $B^6$ and ${\cal S}$ is summarized. Observe that both geometries elegantly avoid  the conditions of the no-go theorem of \cite{papiv}
and so compact examples exist.}
\end{table}

The only remaining condition that needs to be satisfied in order to find a solution for both the Killing spinor and
field equations of the theory is $dH_{(8)}=0$. This leads to
\bea
d\tilde H_{(8)}=k^{-2}dh\wedge dh+k^{-2} d\ell\wedge d\ell+dH_{(6)}=k^{-2} dh\wedge dh+k^{-2} d\ell\wedge d\ell-di_{I_{(6)}} d\omega_{(6)}=0~.
\nn
\eea
One can easily modify this condition if the heterotic anomaly is taken into account. But we shall not investigate this case here.

To summarize, if ${\cal S}$ is taken to be a holomorphic torus fibration over a 6-dimensional
Hermitian manifold, the conditions that must be satisfied to find a solution
are
\bea
&&dh_{ij} \omega^{ij}_{(6)}=0~,~~~d\ell_{ij} \omega^{ij}_{(6)}=-2k^2~,~~~\theta_{\omega_{(6)}}=2d\Phi~,
\cr
&&(\hat\rho_{(6)})_{ij}- d\ell_{ij}=0~, ~~~k^{-2} dh\wedge dh+k^{-2} d\ell\wedge d\ell-di_{I_{(6)}} d\omega_{(6)}=0~.
\la{allcon}
\eea

In what follows, we investigate  these conditions and give some explicit solutions.

\subsection{Reconstruction of spacetime}

The spacetime $M$ of heterotic horizons preserving 4 supersymmetries is a $SL(2, \bR)\times U(1)$ fibration
over $B^6$.  The base space $B^6$ satisfies all the properties mentioned in the previous section
for constructing ${\cal S}$. It remains to give the connection $\lambda$ of $M$. This is expressed in terms of the
connections $h$ and $\ell$ of ${\cal S}$ as follows:
\bea
\lambda^- &=& \bbe^-~,~~~
\lambda^+ = \bbe^+ - {1 \over 2} k^2 u^2 \bbe^- -u h~,~~~
\cr
\lambda^1 &=& k^{-1}\big(h+ k^2 u \bbe^-\big)~,~~~\lambda^6= k^{-1} \ell~,
\la{g2vbi}
\eea
where
\bea
\bbe^-=dr+r h~,~~~ \bbe^+=du~.
\eea
The spacetime metric and torsion are given as
\bea
ds^2&\equiv&2 \lambda^- \lambda^++(\lambda^1)+(\lambda^6)^2+ds_{(6)}^2
\cr
&=&2 du (dr+r h)+ k^{-2} h\otimes h+ k^{-2} \ell\otimes \ell+ ds^2_{(6)}~,
\cr
H&\equiv&CS(\lambda)+H_{(6)}
\cr
&=&du\wedge dr\wedge h+r du\wedge dh+ k^{-2} h\wedge dh+ k^{-2} \ell\wedge d\ell-i_{I_{(6)}} d\omega_{(6)}~.
\la{bhst}
\eea
It is clear that given the geometric data on ${\cal S}$, the geometry of $M$ is completely described.

\subsection{Toric fibrations}

It is clear from the conditions that arise from supersymmetry as well as the examples that we shall investigate
later that toric fibrations are central in the examination of near horizon geometries. Because of this, we shall derive some
useful formulae for the analysis which shall follow. Toric fibrations in the context of manifolds with skew symmetric
torsion and $SU$ holonomy have been investigated before \cite{sethi, gold, grant, yaufu}.

Suppose that $2n$-dimensional manifold  $Y$ is a $T^{2(n-m)}$ fibration over $2m$-dimensional manifold $X$. Write
the metric $ds^2$ and torsion $H$ on $Y$ as
\bea
ds_{(2n)}^2&=&\delta_{ab} \lambda^a \lambda^b+ds_{(2m)}^2~,
\cr
H_{(2n)}&=&\delta_{ab} \lambda^a\wedge d\lambda^b+ H_{(2m)}~,
\eea
where $ds_{(2m)}^2$ and $H_{(2m)}$ are the metric and torsion on $X$.

Suppose now that in addition $X$ is a Hermitian manifold with complex structure $I_{(2m)}$ compatible with $ds_{(2m)}^2$.
As is well known, the condition
\bea
\hat{\tilde{ \nabla}}_{(2m)} I_{(2m)}=0~,
\eea
implies that
\bea
H_{(2m)}=-i_{I_{(2m)}} d\omega_{(2m)}~,
\eea
where $\omega_{(2m)}$ is the Hermitian form of $X$.

In turn, $Y$ admits an almost Hermitian form $\omega_{(2n)}$ compatible with  $ds_{(2n)}^2$ given by
\bea
\omega_{(2n)}=-\sum^{n-m}_{k=1} \lambda^k\wedge \lambda^{n-m+k}+\omega_{(2m)}~.
\eea
The associated almost complex structure $I_{(2n)}$ is integrable   provided that the curvatures
\bea
{\cal F}^a=d\lambda^a
\eea
are (1,1)-forms on $X$,
\bea
({\cal F}^a)^{2,0}=0~.
\eea
 Furthermore, the connection with torsion, $\hat\nabla_{(2n)}$, on $Y$ is compatible with the complex structure, $I_{(2n)}$, on $Y$, ie
\bea
\hat{\nabla}_{(2n)} I_{(2n)}=0~,
\eea
and so
\bea
H_{(2n)}=-i_{I_{(2n)}} d\omega_{(2n)}~.
\eea
In addition\footnote{This has not been observed in \cite{gold}. As a result, the holonomy of the connection with skew-symmetric torsion
of 6-dimensional manifolds which are $T^2$ fibrations is contained in $SU(2)$ rather than $SU(3)$.}
\bea
\hat{\nabla}_{(2n)}\lambda^a=0~.
\eea
As a result, the holonomy of $\hat{\nabla}_{(2n)}$ is contained in $U(m)$,
\bea
{\rm hol}(\hat{\nabla}_{(2n)})\subseteq U(m)~.
\eea

Now let us investigate the conditions under which the holonomy of $\hat{ \nabla}_{(2n)}$ reduces further
to a subgroup of $SU(m)$. For this, we express the curvature of the $\hat{ \nabla}_{(2n)}$ connection in terms of
that of $\hat{ \tilde\nabla}_{(2m)}$ to find

\bea
\hat R_{k\ell,}{}^i{}_j&=& \hat{\tilde R}_{k\ell,}{}^i{}_j- \delta_{ab} {\cal F}^a_{k\ell} {\cal F}^b{}^i{}_j~,
\cr
\hat R_{ab,}{}^i{}_j&=&{\cal F}_a{}^i{}_k {\cal F}_b{}^k{}_j-{\cal F}_b{}^i{}_k {\cal F}_a{}^k{}_j~,
\cr
\hat R_{ak,}{}^i{}_j&=& \hat{\nabla}_k {\cal F}_a{}^i{}_j~.
\eea

A necessary condition\footnote{It is also sufficient if the base space $X$ is simply connected.} for the holonomy of  $\hat{ \nabla}_{(2n)}$ to
reduce to $SU(n)$ is that the Ricci form
\bea
\hat\rho_{(2n)}={1\over4} \hat R_{\uk\uell,}{}^i{}_j I^j{}_i e^\uk\wedge e^\uell~,
\eea
 of $Y$ vanishes. This in turn gives
\bea
 (\hat{\tilde \rho}_{(2m)})_{k\ell}-{1\over2}\delta_{ab}{\cal F}^a_{k\ell}{\cal F}^b{}^i{}_j (I_{(2m)})^j{}_i&=&0 \ ,
 \cr
\hat{\nabla}_k \big({\cal F}_a{}^i{}_j (I_{(2m)})^j{}_i\big)&=&0~.
\eea

The first condition above can be simplified somewhat provided that $X$ is conformally balanced, ie $\theta_{\omega_{(2m)}}=2d\Phi$.
In particular after a bit of computation, one finds that
\bea
\hat{\tilde\rho}_{(2m)}&=&{1\over4}\hat{\tilde R}_{k\ell,}{}^i{}_j (I_{(2m)})^j{}_i \,e^k\wedge e^\ell={1\over2}d d_I \log \det g- 2 d d_I \Phi
\cr
&=&
-i\partial\bar\partial
\log \det g_{(2m)}+4 i\partial\bar\partial\Phi~.
\la{riccif}
\eea
where $\det g_{(2m)}$ is the determinant of the Hermitian metric of $X$, ie $\det g_{(2m)}= \det (g_{(2m)}{}_{\alpha\bar\beta})$.
Observe that the expression for the Ricci form of Hermitian conformally balanced manifolds is very similar to that
of K\"ahler manifolds.

It is important to notice that even if $X$ is conformally balanced,
$Y$ may {\it not} be. In particular, if $X$ is conformally balanced a  sufficient condition for $Y$ to be conformally balanced is
\bea
{\cal F}^a _{ij} \omega^{ij}_{(2m)}=0~.
\eea
We shall use this when we consider  horizon sections which  are $T^4$ fibrations over K\"ahler 4-dimensional manifolds.

\newsection{Horizons with 4 supersymmetries}

\subsection{Forms and geometric conditions}

To solve the conditions (\ref{allcon}), we shall take $B^6$ to be a Hermitian conformally balanced manifold
and the $T^2$-torus fibration to be holomorphic. Write ${\cal S}=P\boxtimes Q$,  where $P$ and $Q$ are principal circle
bundles over $B^6$ associated with the curvatures $dh$ and $d\ell$, respectively.

It is convenient to rewrite the conditions stated in (\ref{allcon}) in form notation. After some straightforward computation,
one finds that

\bea
&&dh\wedge \omega^2_{(6)}=0~,~~~d\ell\wedge \omega^2_{(6)}=-{k^2\over3}\, \omega^3_{(6)}~,~~~d\big(e^{-2\Phi} \omega^2_{(6)}\big)=0~,
\cr
&&\hat\rho_{(6)}- d\ell=0~, ~~~   k^{-2} dh\wedge dh+ k^{-2}d\ell\wedge d\ell-di_{I_{(6)}} d\omega_{(6)}=0~.
\la{allconf}
\eea

The first two conditions can be easily recognized as the Hermitian-Einstein conditions without and with cosmological constant, respectively,
appropriately generalized for Hermitian conformally balanced manifolds. The third condition is the conformal balanced condition. The fourth
condition identifies $Q$ with the canonical bundle $K$ over $B^6$.
Moreover observe that for conformally balanced manifolds
\bea
\hat{\tilde\rho}_{(6)}={1\over2}d d_I \log \det g_{(6)}- 2 d d_I \Phi=
-i\partial\bar\partial
\log \det g_{(6)}+4 i\partial\bar\partial\Phi~,
\eea
where $\det g_{(6)}=\det(g_{(6)}{}_{\alpha\bar\beta})$ is the determinant of the Hermitian metric of $B^6$.

The conditions stated in (\ref{allconf}) can be easily adapted to the cases for which $B^6$ admits some additional structure.
Several possibilities are available. For example, one can assume that $B^6$ with respect to $\omega_{(6)}$ is K\"ahler, conformally K\"ahler
or balanced. In particular, the latter condition requires that the dilaton is constant. We shall demonstrate that
the differential system
(\ref{allconf}) admits solutions. Another class of examples can be generated by taking $B^6$ to be a $T^2$ fibration over a Hermitian 4-manifold. This case will
be investigated separately.   The general case leads to a non-linear system that contains
equations of Monge-Amp\'ere type.

\subsection{Cohomological conditions}

All  conditions in (\ref{allconf}) lead to  restrictions on the cohomology ring of $B^6$.
 To see this observe that the
conformal balanced condition for $B^6$ implies that $e^{-2\Phi} \omega^2_{(6)}$ is closed. Denote the cohomology class
of $e^{-2\Phi} \omega^2_{(6)}$ with $[e^{-2\Phi} \omega^2_{(6)}]$. Then (\ref{allconf}) implies that the cohomology
classes are restricted as
\bea
c_1(P)\wedge [e^{-2\Phi} \omega^2_{(6)}]=0~,~~~c_1(Q)\wedge [e^{-2\Phi} \omega^2_{(6)}]= -{k^2\over6\pi}\, [e^{-2\Phi}\omega^3_{(6)}]~,~~~
\cr
c_1(B^6)-  c_1(Q)=0~,~~~c_1(P)\wedge c_1(P)+c_1(Q)\wedge c_1(Q)=0~,
\la{allconcoh}
\eea
where $c_1$ denotes the first Chern class of the appropriate circle bundle and $c_1(B^6)={1\over 2\pi}[\hat{\tilde\rho}_{(6)}]=K$ is the
first Chern class of the canonical bundle of $B^6$.

The conditions on the cohomology stated above are necessary for the existence of solutions.  This means one should first seek manifolds $B^6$
and $T^2$-bundles over them that satisfy the cohomological conditions (\ref{allconcoh}) and then try to solve the differential conditions (\ref{allconf}).
One can also show that some of the cohomological conditions are also sufficient to find solutions to some of the equations, eg
 solutions for the first equation in (\ref{allconf}). However, it is not known in general that  holomorphic $T^2$ fibrations
over conformally balanced Hermitian manifolds that satisfy all the cohomological conditions (\ref{allconcoh}) are also solutions of (\ref{allconf}).
We shall demonstrate though that (\ref{allconf}) admits solutions by constructing explicit examples.

\subsection{New solution with horizon section $SU(3)$}

We shall demonstrate that ${\cal S}=SU(3)$ is a solution of (\ref{allconf}) with $B^6=SU(3)/T^2$, where $T^2$ is identified
with a maximal torus of $SU(3)$.  The metric of ${\cal{S}}$ is written
\be
d{\tilde{s}}_{(8)}^2 = \delta_{{\underline{i}} {\underline{j}} } e^{\underline{i}} e^{\underline{j}}
\ee
for ${\underline{i}}=1, \dots , 8$, and $e^{\underline{i}}$ satisfy the Maurer-Cartan equations
\be
d e^{\underline{i}} = -{1 \over 2} c_{{\underline{i}}{\underline{j}}{\underline{k}}} e^{\underline{j}} \wedge e^{\underline{k}}
\ee
where $c$ are the structure constants of the Lie algebra  $SU(3)$ with respect to a real basis. In particular, we
work with a normalization with respect to which
\bea
c_{126}&=&c_{135}=c_{368}=-c_{234}=-c_{258}=-c_{456}=-{1 \over \sqrt{2}},
\nn
c_{148}&=&-\sqrt{2}, \qquad
c_{257}=c_{367}=-\sqrt{3 \over 2} \ .
\eea

The hermitian form on $SU(3)$ is taken to be
\be
\omega_{(8)} = e^{\underline{1}} \wedge e^{\underline{4}}+ e^{\underline{2}} \wedge e^{\underline{5}}
 + e^{\underline{3}} \wedge e^{\underline{6}} + e^{\underline{7}} \wedge e^{\underline{8}}~,
\ee
and it is straightforward to check that the Maurer-Cartan equations imply that
$\omega_{(8)}$ is integrable. This complex structure has been introduced on $SU(3)$ in \cite{sevrin} written in a different basis.
The 3-form flux $H_{(8)}=-di_{I_{(8)}} d\omega_{(8)}$
is given by
\be
H_{(8)}= -c~,
\ee
which is covariantly constant with respect to the Levi-civita connection and hence closed,
and $\hat{\tilde \nabla}_{(8)}$ is flat. In the conventions we have adopted, $SU(3)$ can be
considered as a $T^2$ fibration over $B^6$, where the $T^2$ lies in the directions spanned by
$e^{\underline{7}}$ and $e^{\underline{8}}$, so
\be
d{\tilde{s}}_{(6)}^2 = \delta_{ij} e^i e^j, \qquad \omega_{(6)} = e^{1} \wedge e^{4}+ e^{2} \wedge e^{5}
 + e^{3} \wedge e^{6}\ee
for $i,j=1, \dots 6$, $e^i= e^{\underline{i}}$.
Note that the Hermitian form $\omega_{(6)}$ is different from the Kirillov symplectic form put on  regular co-adjoint orbits of semi-simple groups.
In particular, $\omega_{(6)}$ defined above is not closed, however $SU(3)/T^2$ is   balanced since
\bea
d\omega_{(6)}^2=0~,
\eea
and so the dilaton is constant.
The connections $h$, $\ell$ are defined as
\be
h= \sqrt{6} e^{\underline{8}}-\sqrt{2} e^{\underline{7}}, \qquad \ell= -\sqrt{6} e^{\underline{7}}
-\sqrt{2} e^{\underline{8}}
\ee
and so for these solutions, $k=2 \sqrt{2}$. One can use the Maurer-Cartan equations to verify that the associated curvatures $dh$ and $d\ell$
satisfy the conditions (\ref{allconf}), with flux $H_{(6)}=-i_{I_{(6)}} d\omega_{(6)}$
given by
\be
H_{(6)}={1 \over \sqrt{2}} \big( e^1 \wedge e^2 \wedge e^3
+e^1 \wedge e^5 \wedge e^6 - e^2 \wedge e^4 \wedge e^6 - e^3 \wedge e^4 \wedge e^5\big) \ .
\ee
Note that $H_{(6)}$ is not closed. Thus $SU(3)$ satisfies all the properties of a heterotic horizon. The associated
spacetime is $\big(SL(2,\bR)\times SU(3)\big)/U(1)$.

\subsection{Horizon section $S^1\times S^3\times K_3$ }

The $S^1\times S^3\times K_3$  horizon section is a $T^2$ fibration over $B=S^2\times K_3$. There are two ways
of demonstrating this. One is to begin from $S^1\times S^3\times K_3$ and project down on $B=S^2\times K_3$ or alternatively
reconstruct $S^1\times S^3\times K_3$ from $S^2\times K_3$. It is convenient  to do the former operation.
 Of course both procedures give to the same result.
 To begin  introduce the left invariant 1-forms on the 3-spheres as $\sigma^i$,  and  the 1-form $\tau$ along $S^1$. The Maurer-Cartan
 equations give
 \bea
d\sigma^3=\sigma^1\wedge \sigma^2~,~~d\tau=0~,~~~
\la{mc}
\eea
and cyclically in the indices 1,2,3. Then write the metric on ${\cal S}$ as
\bea
ds^2_{(8)}=(\sigma^3)^2+(\sigma^1)^2+(\sigma^2)^2+(\tau)^2+ds^2(K_3)~,
\eea
where $ds^2(K_3)$ is the hyper-K\"ahler metric on $K_3$.

The Hermitian form  on $S^1\times S^3\times K_3$ is
\bea
\omega_{(8)}=-\sigma^3\wedge \tau-\sigma^1\wedge\sigma^2+\omega(K_3)~,
\eea
where $\omega(K_3)$ is a (1,1)-K\"ahler form on $K_3$. It is easy to see that the complex structure is integrable. Moreover, one finds that
\bea
\tilde H_{(8)}=\sigma^1\wedge \sigma^2\wedge \sigma^3~.
\eea
Since the connection with torsion on the group manifold $S^1\times S^3$ is flat, the only contribution in the holonomy of $\hat\nabla_{(8)}$ comes
from the Levi-Civita connection of $K_3$ and so ${\rm hol}(\hat\nabla_{(8)})= SU(2)$. Therefore
\bea
\hat\rho_{(8)}=0~.
\la{8r}
\eea
 Observe though that $S^3\times S^3\times T^2$
is not (conformally) balanced.

To investigate the geometry of $B=S^2\times K_3$ consider the 2-form
\bea
\omega=-\sigma^1\wedge\sigma^2+\omega(K_3)~,
\eea
on $S^1 \times S^3\times K_3$.
If the $T^2$  directions of the fibre are along $\sigma^3$ and $\tau$, $\omega$ descends to a Hermitian form on $B$ as
\bea
i_{\sigma^3}\omega=i_{\tau}\omega=0~,
\eea
and
\bea
{\cal L}_{\sigma^3}\omega=i_{\sigma^3}d\omega=0~,
\eea
and similarly
\bea
{\cal L}_{\tau}\omega=0~.
\eea
So we set $\omega_{(6)}=\omega$. Since $d\omega=0$, $B=S^2\times K_3$ is K\"ahler and so balanced as required.
The dilaton is constant.

The curvatures of the principal bundle connections are
\bea
&&d\ell=\sigma^1\wedge \sigma^2 \ ,
\cr
&&dh=0~.
\eea
In particular the connection which twists $AdS_3$ is flat and so the spacetime is a product $AdS_3\times S^3\times K_3$.
Moreover $d\ell$ is (1,1) and its trace is constant.
Note in addition that $\rho_{(6)}$ does not vanish. It receives a contribution from $S^2$. However, this cancels
the contribution of $d\ell$ so that (\ref{8r}) is satisfied. The solution preserves $1/2$ of the supersymmetry \cite{hh}.

\subsection{New solution with horizon section $S^3\times S^3\times T^2$}
\la{bbs}

The $S^3\times S^3\times T^2$ horizon section is a $T^4$ fibration over $X=S^2\times S^2$. As in the $S^1\times S^3\times K_3$ case investigated
previously, we shall begin from $S^3\times S^3\times T^2$ and project down to $B^6$.
 To begin  introduce the left invariant 1-forms on the two 3-spheres as $\sigma^i$ and $\rho^i$ which satisfy the Maurer-Cartan
 equations
 \bea
d\sigma^3=\sigma^1\wedge \sigma^2~,~~d\rho^3=\rho^1\wedge \rho^2~,~~~
\la{mc2}
\eea
and cyclically in the indices 1,2,3,  respectively. Then write the metric on ${\cal S}$ as
\bea
ds^2_{(8)}=(\sigma^3)^2+(\sigma^1)^2+(\sigma^2)^2+(\rho^3)^2+(\rho^1)^2+(\rho^2)^2
+(\tau^1)^2+(\tau^2)^2~,
\eea
where
\bea
d\tau^1=d\tau^2=0~,
\eea
are the standard 1-forms on $T^2$.

The Hermitian form  on $S^3\times S^3\times T^2$ is
\bea
\omega_{(8)}={1 \over \sqrt{2}} \tau^1\wedge (\sigma^3+\rho^3)-\sigma^1\wedge\sigma^2-\rho^1\wedge \rho^2- {1 \over \sqrt{2}} \tau^2\wedge (\sigma^3-\rho^3)~,
\eea
leading to
\bea
\tilde H_{(8)}=\sigma^1\wedge \sigma^2\wedge \sigma^3+\rho^1\wedge \rho^2\wedge \rho^3~,
\eea
which is the expected 3-form field strength for group manifolds.
It is easy to see using the Maurer-Cartan equations that the associated complex structure is integrable. Moreover
$\hat{\tilde \nabla}_{(8)}$ is flat and so its holonomy is contained in $SU(3)$. Observe though that $S^3\times S^3\times T^2$
is not (conformally) balanced.

The $T^2$ fibration of the horizon section is chosen along the vector fields dual to $\tau^1$ and $\sigma^3+\rho^3$. Moreover set
\bea
\omega=-\sigma^1\wedge\sigma^2-\rho^1\wedge \rho^2- {1 \over \sqrt{2}}  \tau^2\wedge (\sigma^3-\rho^3) \ .
\eea
After a straightforward calculation, it is easy to see that

\bea
i_{\sigma^3+\rho^3}\omega=i_{\tau^1}\omega=0~,
\eea
and
\bea
{\cal L}_{\sigma^3+\rho^3}\omega=i_{\sigma^3+\rho^3}d\omega=0~,
\eea
and similarly
\bea
{\cal L}_{\tau^1}\omega=0~.
\eea
Therefore $\omega$ descends to a Hermitian form on $B^6=(S^3\times S^3)/S^1\times S^1$. So $\omega=\omega_{(6)}$ and the metric on $B^6$ is
\bea
ds^2_{(6)}= {1 \over 2} (\sigma^3-\rho^3)^2+(\sigma^1)^2+(\sigma^2)^2+(\rho^1)^2+(\rho^2)^2
+(\tau^2)^2~.
\eea
Observe that $B$ is balanced
\bea
d\omega_{(6)}^2=0~,
\eea
as expected. Choosing
\bea
h= \sqrt{2} \tau^1~,~~~\ell=\sigma^3+\rho^3~,
\eea
so that $k=\sqrt{2}$,
one finds that
\bea
dh=0~,~~~d\ell=\sigma^1\wedge \sigma^2+\rho^1\wedge \rho^2~.
\eea
So $h, \ell$ satisfy the properties  for the curvatures of the $T^2$ fibration as required by supersymmetry. Since $dh=0$, the spacetime is
 $AdS_3\times S^3\times S^3\times S^1$.
  As we shall demonstrate later, this is the near horizon geometry of two 5-branes intersecting on a string with
the string localized on the transverse space.

\newsection{Horizons as torus fibrations over a 4-manifold}

\subsection{Geometric conditions}
A large class of heterotic horizons can be found by taking $B^6$ to be a $T^2$ fibration over
a 4-dimensional balanced Hermitian manifold $X$.  All 4-dimensional  conformally balanced Hermitian manifolds are conformally K\"ahler.
Since the conformal balanced condition is with respect to the dilaton $\Phi$, one writes
for the Hermitian form of $X$
\bea
\omega_X=e^{2\Phi} \kappa~,~~~d\kappa=0~,
\eea
where $\kappa$ is the K\"ahler form. Therefore ${\cal S}$ is a $T^4$ fibration over $X$.
As before we consider the two principal bundle connections $\ell$ and $h$, set $h=h^1$, and introduce two connections  $h^2$ and $h^3$
along  the two additional torus directions. It follows from  the properties  of torus
fibrations in section 2 that $B^6$ is conformally balanced  provided that
the curvatures $dh^2$ and $dh^3$ are (1,1) and traceless with respect to $\kappa$. Using this,
we summarize the conditions required for a spacetime to preserve 4 supersymmetries as follows
\bea
&&dh^1\wedge \kappa=0~,~~~dh^2\wedge \kappa=0~,~~~dh^3\wedge \kappa=0~,~~~d\ell\wedge \kappa=-{k^2\over2}\, e^{2\Phi} \kappa^2~,~~~d\kappa=0~,
\cr
&&-i\partial\bar\partial \log \det(i\kappa)-d\ell=0~, ~~~
\cr
&&k^{-2}dh^1\wedge dh^1+k^{-2}dh^2\wedge dh^2+k^{-2}dh^3\wedge dh^3+k^{-2} d\ell\wedge d\ell+2 i
\partial\bar\partial e^{2\Phi}\wedge \kappa=0~.
\nn
\la{allconk}
\eea

As in the general case the above conditions lead to restrictions  on the cohomology of $X$. Here, in addition,
the  above differential system can be rewritten as a system of
six equations for six functions. Before we do this, we shall first describe the conditions on the cohomology on $X$.

\subsection{Cohomological conditions}

It is clear that ${\cal S}=P_1\boxtimes P_2\boxtimes P_3\boxtimes Q$, where $P_1,P_2,P_3,Q$ are principal circle bundles over $X$ and $P_1=P$.
The conditions (\ref{allconk}) imply the cohomology
classes are restricted as
\bea
&&c_1(P_1)\wedge [\kappa]=0~,~~~c_1(P_2)\wedge [\kappa]=0~,~~~c_1(P_3)\wedge [\kappa]=0~,~~~
\cr
&&c_1(Q)\wedge [\kappa]= -{k^2\over4\pi}\, [e^{2\Phi}\kappa^2]~,~~~
c_1(X)-c_1(Q)=0~,~~~
\cr
&&c_1(P_1)\wedge c_1(P_1)+c_1(P_2)\wedge c_1(P_2)+c_1(P_3)\wedge c_1(P_3)+c_1(Q)\wedge c_1(Q)=0~,
\la{tallconk}
\eea
where $c_1$ denotes the first Chern class of the appropriate circle bundle and $c_1(X)$ is the
first Chern class of the canonical bundle of $X$.  It is clear from the cohomology conditions above that $c_1(Q)$
must be identified with the {\it canonical} class of $X$. Moreover, the class $c_1(Q)\wedge [\kappa]$ must be a {\it negative}
multiple of the volume class of $X$.

The above conditions can also be rewritten in terms of the intersection form of $X$
\bea
[\alpha]\cdot [\beta]=\int_X \alpha\wedge \beta~.
\eea
The only difference is that the ring product in $H^2(X)$ is replaced with the product as defined by the intersection form.

\subsection{Differential system}

To express the differential conditions as a system of six equations for six functions, we shall use the $\partial\bar\partial$-lemma
and the cohomological conditions stated above. It is assumed that $X$ is chosen such that the conditions
(\ref{tallconk}) have a solution. The procedure resembles the background field method used for quantum calculations in field theory.
Thus one splits the fields, which are represented by the connections, metric, hermitian form and dilaton,  into a background
part and a fluctuation. Though here, the fluctuations are not restricted to be small.
In particular, introduce fixed background fields $\mathring h^1, \mathring h^2, \mathring h^3, \mathring \ell, \mathring \Phi$ and $\mathring \kappa$
which satisfy the cohomological conditions (\ref{tallconk}). Then using the $\partial\bar\partial$-lemma,
there are (1,1)-forms $\alpha^1, \alpha^2, \alpha^3, \chi$ and  $\psi$, and a function $f$, which
depend {\it only} on the background data, such that
\bea
&&d\mathring h^1\wedge \mathring\kappa=i\partial\bar\partial\alpha^1~,~~~d\mathring h^2\wedge \mathring\kappa=i\partial\bar\partial\alpha^2~,~~~
d\mathring h^3\wedge \mathring\kappa=i\partial\bar\partial\alpha^3~,~~~
\cr
&&d\mathring\ell\wedge \mathring\kappa=-{k^2\over2}\, e^{2\mathring\Phi} \mathring\kappa^2+
i\partial\bar\partial\chi~,~~~-i\partial\bar\partial \log \det(i\mathring\kappa)+ i\partial\bar\partial f-d\mathring\ell=0 \ ,
\cr
&&k^{-2}(d\mathring h^1\wedge d\mathring h^1+d\mathring h^2\wedge d\mathring h^2+d\mathring h^3\wedge d\mathring h^3+d\mathring \ell\wedge d\mathring \ell)
+i\partial\bar\partial\psi=0~.
\la{backcon}
\eea

Since we are seeking solutions that preserve the chosen cohomological classes,
and using again the $\partial\bar\partial$-lemma, there are functions $s^1, s^2, s^3, v$ and $w$ such that
\bea
&&dh^1=d\mathring h^1+i\partial\bar\partial s^1~,~~~dh^2=d\mathring h^1+i\partial\bar\partial s^2~,~~~dh^3=d\mathring h^1+i\partial\bar\partial s^3~,
\cr
&&d\ell=d\mathring \ell+i\partial\bar\partial v~,~~~\kappa=\mathring \kappa-i \partial\bar\partial w~,~~~\Phi=\mathring \Phi+\varphi~,
\la{shift}
\eea
where we have also expressed the dilaton\footnote{It is not necessary to split the dilaton in a background and a fluctuation but this was done
 for uniformity.} in terms of the background field and a fluctuation $\varphi$.

Substituting (\ref{shift}) into (\ref{allconk}) and subtracting from the resulting expressions (\ref{backcon}), one finds
\bea
&&i \partial\bar\partial s^1\wedge \mathring \kappa- d\mathring h^1\wedge i \partial\bar\partial w+\partial\bar\partial s^1\wedge\partial\bar\partial w
=-i\partial\bar\partial\alpha^1~,
\cr
&&i \partial\bar\partial s^2\wedge \mathring \kappa-d\mathring h^2\wedge i \partial\bar\partial w+\partial\bar\partial s^2\wedge\partial\bar\partial w
=-i\partial\bar\partial\alpha^2~,
\cr
&&i \partial\bar\partial s^3\wedge \mathring \kappa- d\mathring h^3\wedge i \partial\bar\partial w+\partial\bar\partial s^3\wedge\partial\bar\partial w
=-i\partial\bar\partial\alpha^3~,
\cr
&&i \partial\bar\partial v\wedge \mathring \kappa- d\mathring \ell\wedge i \partial\bar\partial w+\partial\bar\partial v\wedge\partial\bar\partial w=-{k^2\over2}\, e^{2(\mathring \Phi+\varphi)}
(\mathring \kappa-i \partial\bar\partial w)^2+{k^2\over2}\, e^{2\mathring\Phi} \mathring\kappa^2 -i\partial\bar\partial\chi~,
\cr
&&
\det\Big({i\mathring \kappa+\partial\bar\partial w\over i\mathring \kappa}\Big)= e^{-f-v+c}~,
\cr
&&k^{-2}[2i d\mathring h^1\wedge \partial\bar\partial s^1-\partial\bar\partial s^1\wedge \partial\bar\partial s^1+
2i d\mathring h^2\wedge \partial\bar\partial s^2-\partial\bar\partial s^2\wedge \partial\bar\partial s^2+
2i d\mathring h^3\wedge \partial\bar\partial s^3
\cr
&&-\partial\bar\partial s^3\wedge \partial\bar\partial s^3+
2i d\mathring \ell\wedge \partial\bar\partial v-\partial\bar\partial v\wedge \partial\bar\partial v]
+2i \partial\bar\partial e^{2(\mathring \Phi+\varphi)}\wedge (\mathring \kappa-i\partial\bar\partial w) =i\partial\bar\partial\psi~,
\la{nonlin}
\eea
where $c$ is a constant.

This is a non-linear system of six equations for the six functions $s^1, s^2,s^3, v, w$ and $\varphi$. It contains a Monge-Amp\'ere
type of equation. It is easy to see that each equation can be solved for one unknown function treating the
remaining functions as sources. For example if $v$ is a source, the Monge-Amp\'ere equation is  identical to the one solved by Yau
for the proof  of the Calabi conjecture. However, it is less clear that the full non-linear system has solutions.
Therefore the question rises as to whether for every solution of the cohomological conditions (\ref{tallconk}),
there is a smooth solution of (\ref{nonlin}).

There is also the possibility that (\ref{nonlin}) that does not have any solutions at all. This is not the case. We shall
give  explicit examples below which solve all the conditions. Moreover, we shall explore the conditions on the cohomology.
We shall demonstrate that these
have  solutions for $X$ a del Pezzo surface.

\subsection{New solution with horizon section $S^3\times S^3\times T^2$}

The $S^3\times S^3\times T^2$ horizon section is a $T^4$ fibration over $X=S^2\times S^2$. As in the $S^1\times S^3\times K_3$ case investigated
previously, we shall begin from $S^3\times S^3\times T^2$ and project down to $B$ and to $X$.
 To begin  introduce the left invariant 1-forms $\sigma^i$ and $\rho^i$ on the two 3-spheres  which satisfy the Maurer-Cartan
 equations as in (\ref{mc2}).
  Then write the metric on ${\cal S}$ as
\bea
ds^2_{(8)}=(\sigma^3)^2+(\sigma^1)^2+(\sigma^2)^2+(\rho^3)^2+(\rho^1)^2+(\rho^2)^2
+(\tau^1)^2+(\tau^2)^2~,
\eea
where
\bea
d\tau^1=d\tau^2=0~,
\eea
are the standard 1-forms on $T^2$.

The Hermitian form  on $S^3\times S^3\times T^2$ is
\bea
\omega_{(8)}=-\sigma^3\wedge \rho^3-\sigma^1\wedge\sigma^2-\rho^1\wedge \rho^2-\tau^1\wedge \tau^2~,
\eea
leading to
\bea
\tilde H_{(8)}=\sigma^1\wedge \sigma^2\wedge \sigma^3+\rho^1\wedge \rho^2\wedge \rho^3~,
\eea
which is the expected 3-form field strength for group manifolds.
It is easy to see using the Maurer-Cartan equations that the associated complex structure is integrable. Moreover
$\hat{\tilde \nabla}_{(8)}$ is flat and so its holonomy is contained in $SU(3)$. Observe though that $S^3\times S^3\times T^2$
is not (conformally) balanced.

First, we shall investigate the geometry of $B=S^2\times S^2\times T^2$ and then of $X$. For this, consider the 2-form
\bea
\omega=-\sigma^1\wedge\sigma^2-\rho^1\wedge \rho^2-\tau^1\wedge \tau^2~,
\eea
on $S^3\times S^3\times T^2$.
If the fibre directions are along $\sigma^3$ and $\rho^3$, $\omega$ descends to a Hermitian form on $B$ as
\bea
i_{\sigma^3}\omega=i_{\rho^3}\omega=0~,
\eea
and
\bea
{\cal L}_{\sigma^3}\omega=i_{\sigma^3}d\omega=0~,
\eea
and similarly
\bea
{\cal L}_{\rho^3}\omega=0~.
\eea
So we set $\omega_{(6)}=\omega$. Since $d\omega=0$, $B=S^2\times S^2\times T^2$ is K\"ahler and so balanced as required.
The dilaton is constant.

 The canonical bundle of $S^2\times S^2\times T^2$ is not trivial. However, it becomes trivial after pulling it back on
 $S^3\times S^3\times T^2$. To see this consider the (3,0) form
\bea
\chi= {1 \over 2 \sqrt{2}} (\sigma^1+i\sigma^2)\wedge (\rho^1+i\rho^2)\wedge (\tau^1+i\tau^2)~,
\eea
on $S^3\times S^3\times T^2$.
Clearly
\bea
i_{\sigma^3}\chi=i_{\rho^3}\chi=0~,
\eea
but
\bea
{\cal L}_{\sigma^3+\rho^3}\chi=2i \chi~,~~~{\cal L}_{\sigma^3-\rho^3}\chi=0~.
\eea
Since $\chi$ is transformed up to a phase in the $\sigma^3+\rho^3$ direction, the trivial bundle with section $\chi$ over $S^3\times S^3\times T^2$
 is projected down to a non-trivial bundle over $S^2\times S^2\times T^2$.

Furthermore,  one can set
\bea
h= -(\sigma^3-\rho^3)~,~~~\ell= (\sigma^3+\rho^3) \ .
\eea
Then
\bea
dh=-( \sigma^1\wedge \sigma^2-\rho^1\wedge \rho^2)
\eea
which is (1,1) and traceless and
\bea
d\ell=( \sigma^1\wedge \sigma^2+\rho^1\wedge \rho^2)
\eea
which is (1,1) but not traceless. The trace of $d\ell$ is constant as required.

Since $B=S^2\times S^2\times T^2$, the 4-dimensional K\"ahler manifold $X$ is $X=S^2\times S^2$.
The projection from $B$ to $X$ is
along the trivial $T^2$ fibration.
The K\"ahler form on $X$ is $\kappa=-\sigma^1\wedge\sigma^2-\rho^1\wedge \rho^2$.
Clearly, $S^3\times S^3\times T^2$ is a $T^4$ fibration over $S^2\times S^2$
with principal bundle connection $\ell=\ell$, $h^1=h$, $h^2=\tau^1$ and $h^3=\tau^2$. The spacetime is
isomorphic to $(AdS_3\times S^3\times S^3)/S^1\times T^2$.

\subsection{Horizon section $S^3\times T^5$}

To describe the geometry,  we introduce the left invariant 1-forms $\sigma$ on $S^3$ as in the previous example.  We also consider the
1-forms $\tau^3$,  $\tau^4$ and $\tau^5$ spanning the  $T^3$ directions that replace the second $S^3$. So now
\bea
d\tau^3=d\tau^4=d\tau^5=0 \ .
\eea
The analysis is identical to the one presented in the previous example. The only difference is that the 1-forms $\tau$ which replace the
$\rho$'s
are closed instead of satisfying (\ref{mc2}). Thus one finds that $B=S^2\times T^4$, which is K\"ahler, and $X=S^2\times T^2$.

The principal bundle connections are $\ell=\sigma^3$, $h^1=\tau^1$, $h^2=\tau^2$ and $h^3=\tau^3$.
The curvatures of the principal bundle connections are
\bea
d\ell=\sigma^1\wedge \sigma^2~,~~dh^1=dh^2=dh^3=0~.
\eea
Clearly all the curvatures are (1,1) and $d\ell$ has constant trace. The spacetime is $AdS_3\times S^3\times T^4$
and preserves $1/2$ of supersymmetry \cite{hh}. Since $dh=0$, $h=h^1$, $AdS_3$ does not twist over $B$.

\newsection{Solutions of  cohomology conditions}

We have demonstrated that the differential system (\ref{nonlin}) admits solutions. Now we shall provide evidence that
there may be a large class of solutions to (\ref{nonlin}) by proving that there are many
manifolds that satisfy the cohomological conditions (\ref{tallconk}). We shall take $X$ to be $\bP^2$ blown up at $k$
points\footnote{Throughout this section
$k$ is the number of points that $\bP^2$ is blown up and it should not be confused with the normalization factor
that appears in the definition of  metric  and fluxes in previous sections, see eg  (\ref{8data}).}. This surface
has been used before in the context of manifolds with skew-symmetric torsion \cite{grant}. If $k<9$,
then these manifolds are del Pezzo surfaces $dP_{9-k}$. These have found applications in the context of
mysterious duality \cite{vafa}, which relates the U-duality \cite{ht} 1/2 BPS states of M-theory toroidal  compactifications  to rational curves
in del Pezzo surfaces, and supergravity \cite{julia}.

Appropriately choosing the points  that $\bP^2$ is blown up, the cohomology $H^2(X,\bZ)$ is generated by the hypersurface class
$H$ and the exceptional divisors $E_i$, $i=1,\dots, k$ \cite{hart}. The intersection form is
\bea
H\cdot H=1~,~~~H\cdot E_i=0~,~~~E_i\cdot E_j=-\delta_{ij}~.
\la{intrel}
\eea
The anti-canonical  class, $-K$, of $X$ is
\bea
-K=3H-E_1-E_2-\dots-E_k~.
\eea
There is another basis in the cohomology which consists of the anti-canonical class $-K$ and
\bea
&&\alpha_i=E_i-E_{i+1}~,~~~i=1,\dots, k-1~,
\cr
&&\alpha_k=H-E_1-E_2-E_3~.
\la{abasis}
\eea
The anti-canonical class is orthogonal to the rest of the generators. The intersection matrix in the  $(\alpha_i)$, $i=1,\dots, k$, basis is
\bea
K\cdot \alpha_i=0~,~~~\alpha_i\cdot \alpha_j=-A_{ij}~,~~~i,j=1,\dots, k~,
\la{cmatrix}
\eea
where $(A_{ij})$ is the Cartan matrix of exceptional Lie algebras
 $\mathbf{E}_k$, see table 1. As a result, the intersection matrix can be represented with the associated Dynkin diagram.

\begin{table}[ht]
 \begin{center}
\begin{tabular}{|c|c|}
\hline
$k$&$\mathbf{E}_k$ \\
 \hline
 \hline
$1$&$A_1$\\
\hline
$2$&$A_1\oplus A_1$\\
\hline
$3$&$A_2\oplus A_1$\\
\hline
$4$&$A_4$\\
\hline
$5$&$D_5$\\
\hline
$6$&$\mathbf{E}_6$\\
\hline
$7$&$\mathbf{E}_7$\\
\hline
$8$&$\mathbf{E}_8$\\
\hline
$k>9$&$\mathbf{E}_{k}$\\
\hline
\end{tabular}
\end{center}
\caption{\small
The intersection matrix of a del Pezzo surface, $k<9$, is given by the Cartan matrix of exceptional Lie algebras.
 The intersection matrix of of $\bP^2$ blown up at  $k>9$ in general position  are also given by the Cartan matrix of
 exceptional algebras $\mathbf{E}_{k}$. }
\end{table}

The (\ref{tallconk}) conditions require that
\bea
c_1(Q)=K \ .
\la{c1k}
\eea
To solve the remaining conditions in (\ref{tallconk}), one has to choose appropriately the  Chern classes of the bundles $P_1, P_2$ and $P_3$,
and $[\kappa]$.
We shall consider the cases for which $X$ is a del Pezzo surface and for which $k>9$ separately.

\subsection{del Pezzo}

\subsubsection{$k<9$ odd}

There are two cases to consider here depending on the choice of K\"ahler class. Not all classes in $H^2(X, \bR)$ can be represented by a
K\"ahler form. Those that can obey the  Nakai-Moishezon criteria. These will be stated later. In this case,
the most straightforward choice is  $[\kappa]=-K$. Observe that $c_1(Q)\cdot [\kappa]=-K^2=k-9<0$, ie $c_1(Q)\wedge [\kappa]$
is a negative multiple of the volume class of $X$ as required.
Setting
\bea
c_1(P_s)=n_s H- \sum _i m_{si} E_i~,~~~s=1,2,3~,
\eea
we find that the condition $c_1(P_s) \cdot [\kappa]=0$ implies that
\bea
\label{bc1}
3 n_s- \sum _i m_{si}=0~.
\eea
Moreover the cohomological condition which arises from the closure of $H_{(8)}$ reads
\bea
\label{bc2}
9-k+\sum_s \big(n_s^2-\sum_i (m_{si})^2\big)=9-k- \sum_s \big(8 n_s^2+2 \sum_{i<j} m_{si} m_{sj}\big)=0~.
\eea
It is clear that for the above choice of K\"ahler class, there are solutions only if $k$ is odd.
Moreover, it is straightforward to show that for  $k=3, 5, 7$, ({\ref{bc1}}) and ({\ref{bc2}})
imply that $|m_{si}| \leq 2$ for all $s, i$.  So there are only finitely many solutions.
Furthermore one also finds, for $k=3, 5, 7$, that at most one of the $m_{si}$ can take the
value $\pm 2$, the remaining $m_{si}$ must be $\pm 1$ or $0$. One can then enumerate all possible solutions.

\vskip 0.2cm

\underline{$k=1$}

\vskip 0.2cm
The solution for $k=1$ is unique, modulo overall sign changes in the $c_1(P_s)$ and
permutations of the $P_s$. The Chern class  can be chosen, without loss of generality, as
\bea
c_1(P_1)=H-3  E_1~,~~~c_1(P_2)=c_1(P_3)=0 \ .
\eea

\vskip 0.2cm

\underline{$k=3$}

\vskip 0.2cm

There are seven distinct types of solutions up to  permutations in the $P_s$ and overall changes of
sign in $c_1(P_s)$. In particular, one finds
\bea
c_1(P_1) &=& E_{a_1}-E_{a_2}, \quad c_1(P_2)=E_{b_1}-E_{b_2}, \quad c_1(P_3)=E_{f_1}-E_{f_3} \ ,
\nn
c_1(P_1) &=& H-E_1-E_2-E_3, \quad c_1(P_2)= E_{a_1}-E_{a_2}, \quad c_1(P_3)=E_{b_1}-E_{b_2} \ ,
\nn
c_1(P_1) &=& \pm c_1(P_2)= H-E_1-E_2-E_3, \quad c_1(P_3)=E_{a_1}-E_{a_2} \ ,
\nn
c_1(P_1) &=& \pm c_1(P_2) =\pm c_1(P_3) = H-E_1-E_2-E_3 \ ,
\nn
c_1(P_1) &=& -2E_{a_1}+E_{a_2}+E_{a_3}, \quad c_1(P_2)=c_1(P_3)=0 \ ,
\nn
c_1(P_1) &=& H-2E_{a_1}-E_{a_2}, \quad c_1(P_2)=E_{b_1}-E_{b_2}, \quad c_1(P_3)=0 \ ,
\nn
c_1(P_1)&=& H-2 E_{a_1} - E_{a_2}, \quad c_1(P_2) = H-E_1-E_2-E_3, \quad c_1(P_3)=0 \ ,
\eea
where $\{a_i \}$ are distinct, $\{b_i\}$ are distinct, $\{ f_i \}$ are distinct taking values in the set $\{1,2,\dots,k\}$, $k=3$.
Though we do allow for
$a_i=b_j$  and so on.


%

\vskip 0.2cm

\underline{$k=5$}

\vskip 0.2cm

There are seven distinct types of solutions up to permutations in the $P_s$ and overall changes of
sign in $c_1(P_s)$.  In all cases, one can take, without loss of generality
$c_1(P_3)=0$, and the remaining Chern classes $c_1(P_1), c_1(P_2)$ are

\bea
c_1(P_1)&=& -H-E_{a_1}+E_{a_2}+E_{a_3}+E_{a_4}+E_{a_5}, \quad c_1(P_2)=0 \ ,
\nn
c_1(P_1) &=& -E_{a_1}-E_{a_2}+E_{a_3}+E_{a_4}, \quad c_1(P_2)=0 \ ,
\nn
c_1(P_1) &=& -H+E_{a_1}+E_{a_2}+E_{a_3}, \quad c_1(P_2)=E_{b_1}-E_{b_2} \ ,
\nn
c_1(P_1) &=& E_{a_1}-E_{a_2}, \quad c_1(P_2)= E_{b_1}-E_{b_2} \ ,
\nn
c_1(P_1) &=& -H + E_{a_1}+E_{a_2}+E_{a_3}, \quad  c_1(P_2) = -H+E_{b_1}+E_{b_2}+E_{b_3} \ ,
\nn
c_1(P_1) &=& H-2E_{a_1}-E_{a_2}, \quad c_1(P_2)=0 \ ,
\nn
c_1(P_1) &=& 2H-2E_{a_1}-E_{a_2}-E_{a_3}-E_{a_4}-E_{a_5}, \quad c_1(P_2)=0 \ .
\eea

%

\vskip 0.2cm

\underline{$k=7$}

\vskip 0.2cm

There are three distinct types of solutions again up to permutations in the $P_s$ and overall changes of
sign in $c_1(P_s)$. One can take, without loss of generality,
$c_1(P_2)=c_1(P_3)=0$.  The remaining Chern class $c_1(P_1)$ is

\bea
c_1(P_1) &=& -2H+E_{a_1}+E_{a_2}+E_{a_3}+E_{a_4}+E_{a_5}+E_{a_6} \ ,
\nn
c_1(P_1) &=& -H+E_{a_1}+E_{a_2}+E_{a_3} \ ,
\nn
c_1(P_1) &=& E_{a_1}-E_{a_2} \ .
\eea

\subsubsection{$k<9$ even}

As we have mentioned, if the K\"ahler class
is identified with the anti-canonical class $-K$, one cannot solve the cohomological conditions when $k$ is even.
However, this problem can be circumvented by choosing another
K\"ahler class. For this write
\bea
[\kappa]=p H- \sum _{i} q_i E_i~.
\eea
The requirement that $[\kappa]$ is represented by a K\"ahler class restricts the components $p, q_i$. In particular according
to Nakai-Moishezon criteria, $[\kappa]$ must satisfy

\begin{itemize}
\item $[\kappa]\cdot [\kappa]>0$.

\item $[\kappa]\cdot D>0$ for any irreducible curve $D$ with negative self intersection.

\item $[\kappa]\cdot C>0$ for an ample divisor $C$.

\end{itemize}

The divisor class $n H-E_1-E_2\dots -E_k$  is ample for sufficiently large $n$. On the blow up of distinct points on a smooth cubic in $\bP^2$,
the irreducible curves with negative intersection are $E_i$, $H-E_i-E_j$, $i\not=j$ for $k\leq 9$.
The above conditions lead to the restrictions
\bea
p^2>\sum_{i=1}^k q_i^2~,~~~p>q_i+q_j~,~~~q_i> 0~,~~~i=1, \dots, k~.
\la{pqq}
\eea
Writing
\bea
c_1(P_s)= n_s H-\sum_i m_{si} E_i~,
\eea
the conditions $c_1(P_s)\cdot [\kappa]=0$ imply
\bea
n_s p- \sum_i m_{si} q_i=0~.
\eea
In addition, $-c_1(Q)\cdot [\kappa]>0$ gives
\bea
3p-\sum_{i=1}^k q_i>0~.
\la{ppp}
\eea
Furthermore, the condition associated with the closure of $H_{(8)}$
gives
\bea
9-k+\sum_s n_s^2-\sum_s \sum_i (m_{si})^2=0~.
\eea
Examples of solutions to the cohomological conditions are as follows
\vskip 0.2cm

\underline{$k=2$}
\vskip 0.2cm
\bea
[\kappa]= 4H-E_1-E_2~,~~~              c_1(P_1)=H-2E_1-2E_2~,~~~c_1(P_2)=c_1(P_3)=0~.
\eea
\vskip 0.2cm
\underline{$k=4$}
\vskip 0.2cm
\bea
&&[\kappa]= 4H-E_1-E_2-E_3-E_4~,~~~c_1(P_1)=H-E_1-E_2-E_3-E_4~,~~~
\cr
&&c_1(P_2)=E_1-E_2~,~~~c_1(P_3)= 0~.
\eea

\vskip 0.2cm
\underline{$k=6$}
\vskip 0.2cm
\bea
&&[\kappa]= 8H-3E_1-3E_2-3E_3-3E_4-3E_5-3E_6~,~~~c_1(P_2)=c_1(P_3)= 0~,
\cr
&&c_1(P_1)=3H-2E_1-2E_2-E_3-E_4-E_5-E_6~.
\eea
\vskip 0.2cm
\underline{$k=8$}
\vskip 0.2cm
\bea
&&[\kappa]= 17H-6\sum_{i=1}^8 E_i~,~~~c_1(P_2)=c_1(P_3)= 0~,
\cr
&&c_1(P_1)=6H-3E_1-2E_2-2E_3-2E_4-2E_5-2E_6-2E_7-2E_8~.
\eea

A direct observation reveals that $c_1(P_s)^2\leq 0$ as required for classes represented by anti-self-dual 2-forms.
Moreover in all cases $-c_1(Q)\cdot [\kappa]>0$.

It is clear that there are many solutions of the cohomological conditions on del Pezzo surfaces.
So it is likely that the differential system (\ref{allconk}) admits solutions as well.
This will give a new geometry on del Pezzo surfaces as it will be different
from  the Einstein metrics of \cite{tian}. This is because if the Einstein metric is chosen as a solution of the
differential system (\ref{allconk}), then $d\ell$ will be proportional to the K\"ahler form and so the dilaton
has to be constant. As a result, it is not possible to solve the equation which arises from $dH=0$ as the wedge
product of harmonic forms may not be harmonic.

\subsection{$k>9$}

The cohomological system does not have solutions for $k>9$ for any choice of K\"ahler class. This is because
now $K^2=9-k<0$ and  $c^2_1(P_s)\leq 0$ since the latter are represented by anti-self-dual forms. Therefore
the cohomological condition associated with $dH_{(8)}=0$ cannot be satisfied.
This can be confirmed using Nakai-Moishezon criteria
for  choosing the K\"ahler class and the rest of the conditions of the cohomological system. It is worth noting that
for $k\geq 10$, apart from $E_i$, $H-E_i-E_j$, $i\not=j$, $-K$ is also an irreducible curve with negative self-intersection.
Then (\ref{ppp}) arises as part of  the Nakai-Moishezon criteria.

To include solutions with $k>9$, the cohomological condition associated with $d\tilde H_{(8)}=0$ must be modified. One option
is to include contributions from the anomaly cancelation which schematically leads to
\bea
d\tilde H_{(8)}=-{\alpha'\over4} ({\rm tr} R^2-{\rm tr} F^2)~.
\eea
Alternatively, one can add point anti-5-brane sources as
\bea
d\tilde H_{(8)}=n \delta(\bar {NS5})~.
\eea
Choosing these appropriately, one can cancel the negative contributions arising from the  canonical class
and the rest of the Chern classes of the circle bundles.

\subsection{Topology of heterotic horizons}

In all cases, the horizon spacetime is contractible to the horizon section ${\cal S}$.
In turn the horizon sections are $T^2$ fibrations over $B^6$. We shall focus on the case for which $B^6$ is
a $T^2$-fibration over a K\"ahler 4-manifold $X$. Then we shall adapt the calculation for a del Pezzo surface $X=dP_{9-k}$.
We shall compute the cohomology in three different scenarios
depending on the number of non-trivial line bundles which appear in the construction of ${\cal S}$ from $X$.

First suppose that
${\cal S}=T^2\times B$, and   $B$ is a non-trivial $T^2$ fibration over  $X$.
To simplify the computations, let us focus on  de Rham cohomology. To compute the de Rham cohomology of ${\cal S}$, it suffices
to find the cohomology of $B$.  Since $B$ is a non-trivial fibration over $X$, the Chern classes
$c_1(Q)=b_1$ and $c_1(P)=b_2$ can be chosen as the first two basis elements in $H^2(X, \bR)$, ie $H^2(X, \bR)=\bR\langle  b_1,\dots, b_m\rangle$.
 Moreover let $H^*(T^2, \bR)=\bR\langle \theta_1, \theta_2, \theta_1\wedge \theta_2\rangle$.
Using the spectral sequence for a fibration
\bea
E_2^{p,q} =H^p(X, H^q(T^2, \bR))~,~~p=0,\dots, 4~,~~~q=0, \dots, 2~.
\la{e22}
\eea
To find the cohomology of $B$, it suffices to calculate the action of the  $d_2$ differential. This has been done
in appendix A. It turns out that

\bea
H^0(B, \bR)=H^6(B, \bR)=\bR~,~~~H^2(B, \bR)=H^4(B, \bR)=\bR^{m-2}~,~~~H^3(B, \bR)=\bR^{2m-2}~.~~~
\eea

We can also give the intersection matrices of the cohomology of $B$. For this suppose that the intersection matrix  $A$ of $X$ is
\bea
\int_X b_e\wedge b_f=A_{ef}~,~~~e,f=1,\dots,m.
\eea
A basis in the cohomology of $B^6$ is
\bea
H^2(B, \bR)&=&\bR\langle b_a \rangle~,~~a\not=1,2 \ ,
\cr
H^3(B, \bR)&=&\bR\langle \theta_r\wedge b_a, \theta_1\wedge b_1+\theta_2\wedge b_2, \theta_1 \wedge b_2+\theta_2\wedge b_1 \rangle \ .
\eea
Observe that one of the generators of $H^3(B, \bR)$ is the Chern-Simons form.
It is straightforward to compute the integrals
\bea
\int_{B} \omega_{(6)}\wedge b_a\wedge b_b&=&-A_{ab}~,
\cr
\int_B \theta_r \wedge b_a\wedge \theta_s\wedge b_b&=&- \epsilon_{rs} A_{ab} \ ,
\cr
\int_B \theta_r \wedge b_a\wedge (\theta_1\wedge b_1+\theta_2\wedge b_2)&=&
-\epsilon_{r1} A_{a1}-\epsilon_{r2} A_{a2}~,~~~
\cr
\int_B \theta_r \wedge b_a\wedge (\theta_1 \wedge b_2+\theta_2\wedge b_1)&=&
-\epsilon_{r1} A_{a2}-\epsilon_{r2} A_{a1}~,~~~
\cr
\int_B(\theta_1\wedge b_1+\theta_2\wedge b_2)\wedge(\theta_1 \wedge b_2+\theta_2\wedge b_1)&=&-A_{11}+A_{22} \ .
\eea
Apart from the first, the rest give the intersection matrix of $B$.
Note that even though $d\omega_{(6)}\not=0$, the first integral does not depend on the representatives of the classes.

Now suppose that $X=dP_{9-k}$. For the solutions for which  $[\kappa]=-b_1=-K$, one can choose, up to an appropriate rescaling,
$\a=-K$ and $\beta=b_2$ as $K\cdot b_2=0$ and $b_2^2\not=0$ to satisfy the assumptions stated in appendix A to calculate the cohomology.
In the case that $[\kappa]\not=-K$, one chooses
 $\alpha=[\kappa]$. Moreover, there is always a class with the properties of $\beta$.
Clearly, the intersection matrices of $B^6$ inherit the exceptional structure of the cohomology of del Pezzo
surfaces.

Moreover, it turns out that if $X=dP_{9-k}$,  $B^6$ has the same de Rham cohomology as  $(k-1) (S^2\times S^4)\# k (S^3\times S^3)$.
In fact under some additional conditions, it is diffeomorphic to it \cite{grant}. The black holes that can arise from such horizons
have non-trivial topology.

Next suppose that ${\cal S}=S^1\times Y$ and $Y$ is a non-trivial $T^3$ fibration over $X$. In such case, the calculation
explained in appendix A reveals that
\bea
&&H^0(Y, \bR)=H^7(Y, \bR)=\bR~,~~H^2(Y,\bR)=H^5(Y,\bR)=\bR^{m-3}~,~~~
\cr
&&H^3(Y,\bR)=H^4(Y,\bR)=\bR^{3m-4}~.
\eea

Furthermore, if ${\cal S}$ is a non-trivial fibration over $X$, then the calculation in appendix A gives

\bea
&&H^0({\cal S}, \bR)=H^8({\cal S}, \bR)=\bR~,~~H^2({\cal S}, \bR)=H^6({\cal S}, \bR)=\bR^{m-4}~,~~~
\cr
&&
H^3({\cal S}, \bR)=H^5({\cal S}, \bR)=\bR^{4m-7}~,~~~H^4({\cal S}, \bR)=\bR^{6m-8}~.
\eea

It is clear that in all cases the cohomology of the horizon section is non-trivial indicating the existence
of exotic heterotic black holes.

\newsection{Near brane geometries for localized brane intersections}

In all known cases, black hole horizons  also arise as near brane geometries. It is likely
that our heterotic horizon solutions can also be interpreted as the near brane geometry of a brane configuration.
Since the only fluxes that are switched on are the dilaton and the 3-form field strength, it is expected that they
correspond to the near brane geometry of a configuration of 5-branes and fundamental strings. In particular,
we shall show that the $AdS_3\times S^3\times S^3\times S^1$ solution is the near brane geometry of two 5-branes
intersecting on a string with the latter  localized in all transverse directions.
\begin{table}[ht]
 \begin{center}
\begin{tabular}{c c c c c c c c c c c}
$5{\rm brane}$&$0$&$1$& $2$&$3$&$4$&$5$&\\
$5{\rm brane}$&$0$&$1$&$~$&$~$&$~$&$~$&$6$&$7$&$8$&$9$\\
${\rm string}$&$0$&$1$&\\
\end{tabular}
\end{center}
\caption{\small
The worldvolume directions  of the branes are given. }
\end{table}
The metric, 3-form field strength and dilaton for this configuration can be written as
\bea
ds^2&=&h^{-1}_s(x,y) (-dt^2+d\sigma^2)+h_5(x) dx^2+h'_5(y) dy^2~,
\cr
H&=&dt\wedge d\sigma\wedge dh_s^{-1}+\star_x dh_5+ \star_y d h'_5~,
\cr
e^{2\Phi}&=&h_s^{-1}h_5h'_5~,
\eea
where $dx^2$ and $dy^2$ is the Euclidean metric on the transverse spaces $\bR^4$ of the corresponding 5-brane and
all the Hodge duals have been taken with the flat metric. If
\bea
h_5=1+{q_5\over |x|^2}~,~~~h'_5=1+{q'_5\over |y|^2}~,
\eea
then $dH=0$. To determine $h_s$, one has to solve the field equation for the 2-form gauge potential which gives
\bea
h_5'\delta^{ab}\partial_a  \partial_b h_s+h_5 \delta^{a'b'}\partial_{a'} \partial_{b'} h_s=0~.
\eea
A solution for $h_s$  is
\bea
h_s=1+{q_1\over |x|^2}+{q'_1\over |y|^2}+{s\over |y|^2 |x|^2}~,
\eea
see also \cite{khuri, sken}.
The geometry near the common intersection is recovered in the limit $|y|^2,  |x|^2\rightarrow 0$ with
the ratio $|y|/|x|$ fixed. In this limit, the last term in the above equation dominates.
Evaluating the metric near this limit after a change of co-ordinates one finds
$AdS_3\times S^3\times S^3\times S^1$ with constant dilaton. Moreover
\bea
H=2 d{\rm vol}(AdS_3)+ d{\rm vol}(S^3)+d{\rm vol}(S^3)~.
\eea
This is the near horizon example which we have found in section \ref{bbs}.

\newsection{ Near horizon geometries for type II black holes}

The solutions we have found have closed 3-form field strength and so they are solutions of type IIA and  IIB supergravities.
As solutions to IIA supergravity, they can be lifted to 11 dimensions. In particular the general form of the solution
in (\ref{bhst}) lifted to 11 dimensions reads
\bea
ds^2_{(11)}&=& e^{-{2\Phi\over3}} ds^2_{(10)}+ e^{4\Phi\over3} dy^2
\cr
F&=& dy\wedge H \ .
\eea
Reducing the solution back to IIA along the non-trivial fibre direction $h$, we find that
\bea
ds^2_{(A)}&=& e^{-\Phi} [-k^2 r^2du^2+2du dr+ k^{-2} \ell\otimes \ell+ds^2_{(6)}+ e^{4\Phi\over3} dy^2]
\cr
F_2&=&k^{-2}(dh+dr\wedge du)~,~~~
H=-k^{-2} dy\wedge dh~,
\cr
F_4&=& dy\wedge[  k^{-2} \ell\wedge d\ell+H_{(6)}]~,~~~e^{2\Phi_{(A)}}=e^{-\Phi}~.
\eea
One may notice that the metric $e^{\Phi} ds^2_{(10)}$ has an $AdS_2$ spanned by the coordinates $r,u$.
Therefore if the dilaton is constant, and this is the case for many of our solutions, the resulting spacetime
can be interpreted as the near horizon geometry of IIA black holes. The solution preserves all 4 supersymmetries as
the reduction is done along $h$ and $dh$ is (1,1) and traceless. This is because  the spinorial Lie derivative of all the Killing spinors
along  the vector
field associated to $h$  vanishes \cite{gpt}.

The solution can be further be T-dualized to IIB along the direction of  $\ell$. In this case, the metric of the
IIB background is
\bea
ds^2_{(B)}&=&e^{-\Phi} [-k^2 r^2du^2+2du dr+ ds^2_{(6)}+ e^{4\Phi\over3} dy^2]+ k^2e^\Phi dx^2 \ .
\eea
Again for constant dilaton, the metric has an $AdS_2$ component. Therefore such spacetimes can be interpreted
as the near horizon geometries of IIB black holes. However in this case, all 4 supersymmetries of the IIA background will be
broken after T-duality to IIB as $d\ell$ although (1,1) is not traceless. The spinorial Lie derivative along  the vector field
associated to $\ell$ on the Killing spinor will not vanish.

Supersymmetry under T-duality can be preserved provided that $B^6$ satisfies additional properties and the T-duality
operation is taken along another direction. In particular if $B^6$ is a $T^2$- fibration over a K\"ahler manifold, then
the two additional fibre directions are associated with curvatures $dh^r$, $r=2,3$, which are (1,1) and traceless. In this case,
after T-duality along any of these two directions one obtains a IIB background with an $AdS_2$ factor
preserving 4 supersymmetries. This is because again the spinorial Lie derivative of the IIA Killing spinor vanishes
when is taken along one of the two fibre directions. So even though IIB supergravity does not
have a 1-form gauge potential, there are supersymmetric black hole near horizon geometries.

\newsection{From horizons to black holes}

Having found a near horizon geometry, it is natural to ask  whether this can be extended to a full black hole
geometry. It is not expected
that all near horizon geometries will give rise to black hole solutions. In the absence of a  solution
to  the field equations,  some qualitative tests have been devised.  One such test is based on the
expectation that there is a continuous interpolation of the horizon section to the compact section of the asymptotic geometry via
a Cauchy surface which lies outside the horizon. Such a test suggests that the horizon and asymptotic sections are {\it cobordant}.
The cobordism mostly used for this is the oriented cobordism ring $\Omega_*$, see \cite{kl}. However, it seems to us that in the presence of spinors which
must be defined at both the horizon and the asymptotic region of spacetime,
and in particular supersymmetry, the most relevant equivalence is that of spin cobordism ring $\Omega^{\mathfrak{spin}}_*$.
$\Omega^{\mathfrak{spin}}_n$, $n\leq 8$, has been computed in \cite{milnor}, see also references within. Here we shall use that
\bea
\Omega^{\mathfrak{spin}}_1=\Omega^{\mathfrak{spin}}_2=\bZ_2~,~~~\Omega^{\mathfrak{spin}}_4=\bZ~,~~~\Omega^{\mathfrak{spin}}_3=\Omega^{\mathfrak{spin}}_5=
\Omega^{\mathfrak{spin}}_6=\Omega^{\mathfrak{spin}}_7=0~,~~~\Omega^{\mathfrak{spin}}_8=\bZ\oplus \bZ~.
\eea
Moreover $\Omega^{\mathfrak{spin}}_1$ is generated by the circle with the periodic spin structure,
and $\Omega^{\mathfrak{spin}}_8$ is generated by $\bP^2(\bH)$, ie the space of quaternionic lines in $\bH^3$, and a manifold $L^8$
satisfying the relation $4 L^8=K_3\times K_3$. Clearly  heterotic horizons lie in $\Omega^{\mathfrak{spin}}_8$.

These data indicate that all the solutions\footnote{All our horizon sections  admit metrics
with positive scalar curvature and so satisfy the positive Yamabe type condition of \cite{schoen}.}
we have found, including a large class  which may arise
from del Pezzo surfaces, can be the near horizon geometries of asymptotically flat or asymptotically AdS black holes.
For such black holes the asymptotic section is a sphere $S^8$ and so it represents
the trivial class in $\Omega^{\mathfrak{spin}}_8$. This is also the case for most Kaluza-Klein black holes.
On the other hand  all our explicit horizon sections, apart from $SU(3)$, are products $S^3\times Z$. Since
 $S^3$ represents the trivial class, because $\Omega^{\mathfrak{spin}}_3$ vanishes, one can consider $D^4\times Z$ which
 has boundary $S^3\times Z$ and so $S^3\times Z$ also represents the trivial class in $\Omega^{\mathfrak{spin}}_8$.
 Therefore all such horizon sections can be associated with asymptotically flat or AdS black holes.
 The same argument applies for the examples based on del Pezzo surfaces since for most of them
 the horizon section is ${\cal S}=S^1\times M^7$. Since $\Omega^{\mathfrak{spin}}_7=0$, one can again argue
  that $S^1\times M^7$ represents the trivial class in $\Omega^{\mathfrak{spin}}_8$.

It may appear that the cobordism equivalence between the horizon and asymptotic sections do not impose much restriction.
This is mostly the case but not always. One of the near horizon geometries that arises in heterotic theory has
section $K_3\times K_3$ \cite{hh}. As we have mentioned this represents a non-trivial class in $\Omega^{\mathfrak{spin}}_8$.
Therefore it is not in the same cobordism class as $S^8$ and so it cannot be the near horizon geometry of an asymptotically
flat or AdS black hole. Moreover, it cannot be the near horizon geometry of most Kaluza-Klein black holes which arise from lifting a 4- or 5- dimensional
black hole to 10 dimensions. This is because
the asymptotic section of such Kaluza-Klein black holes is expected to be either products of spheres or $S\times H$, where again $S$ is a product of spheres
and $H$ is a special holonomy manifold, ie $H$ is either $K_3$ or a 6-dimensional Calabi-Yau. However
such spaces  represent the trivial class in $\Omega^{\mathfrak{spin}}_8$ and so cannot be cobordant to $K_3\times K_3$.

\newsection{Concluding Remarks}

We have constructed explicit examples of near horizon geometries of heterotic supergravity which
preserve 4 supersymmetries. Amongst the horizon sections we have found are $SU(3)$ and $S^3\times S^3\times T^2$. The near horizon
geometry in the $SU(3)$ case is $(SL(2,\bR)\times SU(3))/U(1)$ and $SL(2,\bR)$ is twisted with respect to a $U(1)$ connection.
The horizon section $S^3\times S^3\times T^2$ gives rise to two different near horizon geometries. One
near horizon geometry is $AdS_3\times S^3\times S^3\times S^1$. But there is also the possibility that the near horizon geometry
is $(AdS_3\times S^3\times S^3)/S^1\times T^2$. Therefore a horizon section does {\it not} determine the near horizon
geometry uniquely. To our knowledge, it is the first time that such a possibility has been observed.

We have also demonstrated that a large class of solutions can arise
provided that the horizon section is chosen to be a $T^4$ fibration over 4-dimensional a K\"ahler manifold $X$.
The resulting differential system contains 6 equations for 6 unknown functions which include the Monge-Amp\'ere
equation and a conformally rescaled Hermitian-Einstein equation. We have shown  that the non-linear system has solutions
for $X=\bP^1\times \bP^1$. We have also given a set of  conditions in the cohomology of $X$ which are necessary for
the existence of solutions. We have found many solutions of this cohomological system
when  $X$ is a del Pezzo surface. We have also raised the question of whether for every solution
of the cohomological system there is a solution of the differential equations and so a new heterotic horizon.

We have investigated the topology of the horizons we have found by computing their de Rham cohomology.
Those horizons that are associated with del Pezzo surfaces exhibit cohomology with intersection matrices
which are closely related to exceptional groups. Of course this is inherited from the intersection matrix of del Pezzo
surfaces which is the Cartan matrix of exceptional Lie algebras. The cohomological properties of our horizons
point to a relation to U-duality invariant brane configurations but we have not been able to make this more precise.

There is not an apriori  reason to believe that a  near horizon solution is associated with a black hole.
However, there are some tests that one can do. One of them is to argue that the asymptotic section of a black hole is
cobordant to the horizon section.
In the presence of spinors, this means that both sections  are spin cobordant.
We have found that all our horizon sections which preserve 4 supersymmetries are in the trivial class and so
they may be near horizons of asymptotically flat or AdS black holes.

The plethora of near horizon solutions we have found  suggest that  there are many exotic supersymmetric black holes in heterotic supergravity.
Some of them will be Kaluza-Klein black holes but there is a  possibility that some will have a purely 10-dimensional origin.
The existence of such black holes  will give new insights into  string theory and M-theory.

\vskip 0.5cm
{\bf Acknowledgments:}~We would like to thank Geo Grantcharov
for correspondence and Fabio Riccioni for helpful
discussions. JG is supported by the EPSRC grant EP/F069774/1. GP is
 partially supported
by the EPSRC grant EP/F069774/1 and the STFC rolling grant ST/G000/395/1.

\setcounter{section}{0}
\setcounter{subsection}{0}

\appendix{Topology of horizon sections}

To compute the cohomology of the total space of the torus fibration, we shall use the spectral sequences method
explained in \cite{bott}, see also \cite{grant}.
Suppose that ${\cal S}=T^2\times B$, where $B$ is a non-trivial $T^2$ fibration over $X$. $X$ is simply connected. In such a case, the $E_2^{*,*}$ part of
the spectral sequence defined in (\ref{e22}) is summarized in table 4.

\begin{table}[ht]
 \begin{center}
\begin{tabular}{|c|c|c|c|c|}
\hline
$\bR\langle\theta_{12}\rangle$&$0$  & $\bR\langle \theta_{12}\wedge b_a\rangle$ & $0$&$\bR\langle \theta_{12}\wedge v\rangle$\\
 \hline
$\bR\langle\theta_r\rangle$&$0$&$\bR\langle \theta_r\wedge b_a\rangle$ &$0$&$\bR\langle \theta_r\wedge v\rangle$\\
\hline
$\bR$&$0$&$\bR\langle b_a\rangle$&$0$&$\bR\langle v\rangle$\\
\hline
\end{tabular}
\end{center}
\caption{\small
The entries are the elements of of $E^{*,*}_2$}
\end{table}

To compute the action of the $d_2$ differential first observe that $d_2:\bR\langle\theta_r \rangle\rightarrow \bR\langle b_a\rangle$ with $d_2(\theta_r)=b_r$ which is clearly 1-1. Next
 $d_2:\bR\langle\theta_{12} \rangle\rightarrow \bR\langle\theta_r\wedge b_a\rangle$ with $d_2(\theta_{12})=\theta_1\wedge b_2-\theta_2\wedge b_1$
 which is also 1-1. Moreover  $d_2:\bR\langle\theta_r\wedge b_a \rangle\rightarrow \bR\langle v\rangle$ with $d_2(\theta_r\wedge b_a)=b_r\wedge b_a$,
 where $v$ is the volume class of $X$.
Suppose now that there are  $\a,\b\in H^2(X, \bR)$ such that
\bea
b_1\wedge \alpha=v~,~~~b_2\wedge \beta=v~,~~~b_1\wedge \beta=b_2\wedge \alpha=0~.
\eea
In such a case, $d_2:\bR\langle\theta_r\wedge b_a \rangle\rightarrow \bR\langle v\rangle$ is onto.
Furthermore, $d_2: \bR\langle \theta_{12}\wedge b_a\rangle\rightarrow \bR\langle \theta_r\wedge v\rangle$ with
$d_2(\theta_{12}\wedge b_a)=\theta_1\wedge b_2\wedge b_a-\theta_2\wedge b_1\wedge b_a$ is also onto.
As a result, $E_3^{*,*}$ is given in table 5 and converges in the cohomology of the bundle space.

\begin{table}[ht]
 \begin{center}
\begin{tabular}{|c|c|c|c|c|}
\hline
$0$&$0$  & $\bR^{m-2}$ & $0$&$\bR$\\
 \hline
$0$&$0$&$\bR^{2m-2}$ &$0$&$0$\\
\hline
$\bR$&$0$&$\bR^{m-2}$&$0$&$0$\\
\hline
\end{tabular}
\end{center}
\label{tab1}
\caption{\small
The entries are the elements of of $E^{*,*}_3$}
\end{table}

Therefore the cohomology of $B$ is given by
\bea
H^0(B, \bR)=H^6(B, \bR)=\bR~,~~~H^2(B, \bR)=H^4(B, \bR)=\bR^{m-2}~,~~~H^3(B, \bR)=\bR^{2m-2}~.~~~
\eea

Suppose that ${\cal S}=S^1\times Y$, where $Y$ is a non-trivial $T^3$ fibration over $X$. To find the cohomology of $Y$, the $E_2^{*,*}$ part of the
the spectral sequence is given in table 6.
\begin{table}[ht]
 \begin{center}
\begin{tabular}{|c|c|c|c|c|}
\hline
$\bR\langle\theta_{123}\rangle$&$0$  & $\bR\langle \theta_{123}\wedge b_a\rangle$ & $0$&$\bR\langle \theta_{123}\wedge v\rangle$\\
\hline
$\bR\langle\theta_{rs}\rangle$&$0$  & $\bR\langle \theta_{rs}\wedge b_a\rangle$ & $0$&$\bR\langle \theta_{rs}\wedge v\rangle$\\
 \hline
$\bR\langle\theta_r\rangle$&$0$&$\bR\langle \theta_r\wedge b_a\rangle$ &$0$&$\bR\langle \theta_r\wedge v\rangle$\\
\hline
$\bR$&$0$&$\bR\langle b_a\rangle$&$0$&$\bR\langle v\rangle$\\
\hline
\end{tabular}
\end{center}
\label{ttt}
\caption{\small
The entries are the elements  of $E^{*,*}_2$}
\end{table}
Moreover observe that $d_2:\bR\langle\theta_r \rangle\rightarrow \bR\langle b_a\rangle$ with $d_2(\theta_r)=b_r$ which is clearly 1-1. Next
 $d_2:\bR\langle\theta_{rs} \rangle\rightarrow \bR\langle\theta_r\wedge b_a\rangle$ with $d_2(\theta_{rs})=\theta_r\wedge b_s-\theta_s\wedge b_r$
 which is also 1-1. Moreover  $d_2:\bR\langle\theta_r\wedge b_a \rangle\rightarrow \bR\langle v\rangle$ with $d_2(\theta_r\wedge b_a)=b_r\wedge b_a$.
Suppose now that there are  $\a_s\in H^2(X, \bR)$ such that
\bea
b_r\wedge \alpha_s=\delta_{rs} v~.
\eea
In such a case, $d_2:\bR\langle\theta_r\wedge b_a \rangle\rightarrow \bR\langle v\rangle$ is onto.
Furthermore, $d_2: \bR\langle \theta_{rs}\wedge b_a\rangle\rightarrow \bR\langle \theta_r\wedge v\rangle$ with
$d_2(\theta_{rs}\wedge b_a)=\theta_r\wedge b_s\wedge b_a-\theta_s\wedge b_r\wedge b_a$ is also onto. Next $d_2: \bR\langle\theta_{123}\rangle\rightarrow
\bR\langle \theta_{rs}\wedge b_a\rangle$ with $d_2 (\theta_{123})=\theta_{12}\wedge b_3+\theta_{31}\wedge b_2+\theta_{23}\wedge b_1$, and
$d_2: \bR\langle \theta_{rs}\wedge b_a\rangle\rightarrow \bR\langle \theta_r\wedge v\rangle$ with
$d_2(\theta_{rs}\wedge b_a)=\theta_r\wedge b_s\wedge b_a-\theta_s\wedge b_r\wedge b_a$. The former map is $1-1$ and the second is onto.
As a result, $E_3^{*,*}$ is given in table 7 and converges in the cohomology of the bundle space.

\begin{table}[ht]
 \begin{center}
\begin{tabular}{|c|c|c|c|c|}
\hline
$0$&$0$  & $\bR^{m-3}$ & $0$&$\bR$\\
\hline
$0$&$0$  & $\bR^{3m-4}$ & $0$&$0$\\
 \hline
$0$&$0$&$\bR^{3m-4}$ &$0$&$0$\\
\hline
$\bR$&$0$&$\bR^{m-3}$&$0$&$0$\\
\hline
\end{tabular}
\end{center}
\caption{\small
The entries are the elements of of $E^{*,*}_3$}
\end{table}

In particular, ones finds that
\bea
&&H^0(Y, \bR)=H^7(Y, \bR)=\bR~,~~H^2(Y,\bR)=H^5(Y,\bR)=\bR^{m-3}~,~~~
\cr
&&H^3(Y,\bR)=H^4(Y,\bR)=\bR^{3m-4}~.
\eea

Suppose that ${\cal S}$ is a non-trivial $T^4$ fibration over $X$. The $E_2^{*,*}$ of part of the spectral sequence is
given in table 8.
\begin{table}[ht]
 \begin{center}
\begin{tabular}{|c|c|c|c|c|}
\hline
$\bR\langle\theta_{1234}\rangle$&$0$  & $\bR\langle \theta_{1234}\wedge b_a\rangle$ & $0$&$\bR\langle \theta_{1234}\wedge v\rangle$\\
\hline
$\bR\langle\theta_{rst}\rangle$&$0$  & $\bR\langle \theta_{rst}\wedge b_a\rangle$ & $0$&$\bR\langle \theta_{rst}\wedge v\rangle$\\
\hline
$\bR\langle\theta_{rs}\rangle$&$0$  & $\bR\langle \theta_{rs}\wedge b_a\rangle$ & $0$&$\bR\langle \theta_{rs}\wedge v\rangle$\\
 \hline
$\bR\langle\theta_r\rangle$&$0$&$\bR\langle \theta_r\wedge b_a\rangle$ &$0$&$\bR\langle \theta_r\wedge v\rangle$\\
\hline
$\bR$&$0$&$\bR\langle b_a\rangle$&$0$&$\bR\langle v\rangle$\\
\hline
\end{tabular}
\end{center}
\caption{\small
The entries are the elements of  $E^{*,*}_2$}
\end{table}

A similar analysis to the one we have presented for the two similar cases above gives that the elements of $E_3^{*,*}$ are
given in table 9.

\begin{table}[ht]
 \begin{center}
\begin{tabular}{|c|c|c|c|c|}
\hline
$0$&$0$  & $\bR^{m-4}$ & $0$&$\bR$\\
\hline
$0$&$0$  & $\bR^{4m-7}$ & $0$&$0$\\
\hline
$0$&$0$  & $\bR^{6m-8}$ & $0$&$0$\\
 \hline
$0$&$0$&$\bR^{4m-7}$ &$0$&$0$\\
\hline
$\bR$&$0$&$\bR^{m-4}$&$0$&$0$\\
\hline
\end{tabular}
\end{center}
\caption{\small
The entries are the elements of of $E^{*,*}_3$}
\end{table}

In particular, one finds that
\bea
&&H^0({\cal S}, \bR)=H^8({\cal S}, \bR)=\bR~,~~H^2({\cal S}, \bR)=H^6({\cal S}, \bR)=\bR^{m-4}~,~~~
\cr
&&
H^3({\cal S}, \bR)=H^5({\cal S}, \bR)=\bR^{4m-7}~,~~~H^4({\cal S}, \bR)=\bR^{6m-8}~.
\eea

\appendix{{ A generalization of the differential system}}

The differential (\ref{allconk}) and cohomological (\ref{tallconk}) systems can be easily generalized to holomorphic $T^{2n}$-fibrations
$Y$ over 4-dimensional K\"ahler manifolds $X$.  First consider a conformal scaling of the K\"ahler form $\kappa$ of $X$ as
\bea
\omega_X=e^{2\Phi} \kappa~,~~~d\kappa=0~.
\eea
Then introduce connections $h^r$ adapted to the $T^n$ fibration for which the curvature $dh^r$ is (1,1). The metric and 3-form field
strength on $Y$ are
\bea
ds^2=G_{rs} h^r h^s+ e^{2\Phi} ds^2(X)
\cr
H=G_{rs} h^r\wedge dh^s-i_I de^{2\Phi}\wedge \kappa
\eea
where $G_{rs}$ is a constant fibre metric and $I$ is the complex structure on $X$. The Hermitian form   on $Y$ is
\bea
\omega(Y)= \phi_{rs} h^r\wedge h^s+ e^{2\Phi} \kappa
\eea
where $\phi$ is a compatible constant  Hermitian form which together with $G$ gives rise to a complex structure on the fibre.
Requiring that the connection $\hat\nabla$ on $Y$ has holonomy contained in $SU(2)\subset SU(2+n)$, one finds that
\bea
&&dh^r\wedge \kappa=  {v^r\over2}\, e^{2\Phi} \kappa^2
\cr
&&-i\partial\bar\partial \log \det(i\kappa)+2 i\partial\bar\partial\Phi+ G_{rs}v^r dh^s=0~, ~~~
\cr
&&G_{rs} dh^r\wedge dh^s+2 i
\partial\bar\partial e^{2\Phi}\wedge \kappa=0~,
\la{allconka}
\eea
where $v^r$ are constants. One can rewrite the differential equations in terms of a non-linear system for $2n+2$ functions.
The calculation is an adaptation of the one we have done already for heterotic horizons and we shall not repeat it here.

As for heterotic horizons, there is an associated cohomological system given by

\bea
&&c_1(P_r)\wedge [\kappa]={1\over2}\, [e^{2\Phi}\kappa^2]~,~~~
c_1(X)+v_sc_1(P^s)=0~,~~~
\cr
&& G_{rs} c_1(P^r)\wedge c_1(P^s)=0~,
\la{tallconka}
\eea
where $c_1(P^r)$ denotes the first Chern class of the $P^r$ circle bundle,   $c_1(X)$ is the
first Chern class of the canonical bundle of $X$ and $v_s=G_{rs}v^r$. Since  $c_1(P^r),c_1(X)\in H^2(X,\bZ)$, it is required that
$G_{rs}v^r c_1(P^s)\in H^2(X,\bZ)$ imposing restrictions on the fibre metric and $v$.


\end{document}